\documentstyle [floats,preprint,aps,fixes,epsfig]{revtex}

\newcommand{\half}{{\textstyle {1 \over 2}}}
\newcommand{\qtr}{{\textstyle {1 \over 4}}}
\newcommand{\e}{\mbox{e}}
\newcommand{\tr}{\mbox{tr}}

\def\tr{{\rm tr}\,}
\def\id{{\bf 1}}
\def\Veff{V_{\rm eff}}
\def\Ueff{U_{\rm eff}}
\def\V{\Omega}

\ifpreprintsty
    \advance\textwidth 0.20in
    \advance\hoffset  -0.10in
\fi

\begin{document}
\setcounter{page}{0}
\preprint {\vbox{\baselineskip14pt\hbox{UW/PT-00-12}\hbox{UCLA/00/TEP/21}}}

\title {Large $N$ Quantum Time Evolution Beyond Leading Order}
\author {Anton V.~Ryzhov\footnote {\tt ryzhovav@physics.ucla.edu}}
\address
    {%
    Department of Physics and Astronomy, UCLA, Los Angelos,
    California 90095-1547
    }
\author {Laurence G.~Yaffe\footnote {\tt yaffe@phys.washington.edu}}
\address
    {%
    Department of Physics,
    University of Washington,
    Seattle, Washington 98195-1560
    }%
\date{\today}
\maketitle

\begin {abstract}%
    For quantum theories with a classical limit
    (which includes the large $N$ limits of typical field theories),
    we derive a hierarchy of evolution equations for equal time
    correlators which systematically incorporate corrections
    to the limiting classical evolution.
    Explicit expressions are given for next-to-leading order,
    and next-to-next-to-leading order time evolution.
    The large $N$ limit of $N$-component vector models,
    and the usual semiclassical limit of point particle quantum mechanics
    are used as concrete examples.
    Our formulation directly exploits the appropriate group structure
    which underlies the construction of suitable coherent states and
    generates the classical phase space.
    We discuss the growth of truncation error with time,
    and argue that truncations of the large-$N$ evolution equations
    are generically expected to be useful only for times short compared
    to a ``decoherence'' time which scales like $N^{1/2}$.
\end {abstract}

\thispagestyle {empty}
\vfill
\pagebreak

\section{Introduction}\label{introduction}

The time evolution of quantum systems away from equilibrium is of interest
in many applications including, but certainly not limited to,
phase transition dynamics, inflationary reheating, and
heavy ion collisions.
Large $N$ expansions have provided a widely used technique
for studying equilibrium properties in statistical physics
and field theory \cite {amit,zinn-justin,Coleman},
and it is natural to apply a similar strategy for studying
non-equilibrium problems.
The large $N$ limit (as typically formulated) is actually a special
type of classical limit \cite {Yaffe}.
Suitable observables behave classically and the quantum dynamics
reduces to classical dynamics on an appropriate phase space.

Considerable work has been done
examining the dynamics of far from equilibrium states
in a variety of applications
using leading large-$N$ time-evolution
\cite {Cooper1,Cooper2,Boyanovsky1,Boyanovsky2,Boyanovsky3,Others}.
A major virtue of large $N$ techniques
(compared to alternative wholly uncontrolled approximation schemes)
is that one should be able to improve
the approximation by systematically including sub-leading
effects suppressed by powers of $1/N$.
For a variety of equilibrium problems (such as critical phenomena),
this approach can work quite well
\cite {critical-exponents-1,critical-exponents-2,bose-cond}.

For initial value problems, in which one would like to choose
a non-equilibrium initial state and then examine the subsequent
time evolution, traditional formulations of large-$N$ expansions
using graphical or functional integral techniques \cite {Coleman}
are very awkward.
A major difficulty with these approaches is that they generate
integral equations which are non-local in time when sub-leading $1/N$
corrections are retained.
For practical (numerical) applications, one would vastly prefer
a formulation in which locality in time is always preserved.

In this paper, we describe a formulation of large $N$
(or semi-classical) dynamics which leads to a coupled hierarchy
of time-local evolution equations for equal time correlation
functions.
Our approach directly exploits the appropriate group structure
underlying the construction of suitable coherent states
and the existence of the classical limit \cite {Yaffe}.
We specifically focus on the time evolution of initial states
chosen to equal one of these coherent states. 
We will give explicit next-to-leading order (NLO),
and next-to-next-to-leading order (NNLO), expressions
for the required evolution equations.
Somewhat related hierarchies of evolution equations
have been discussed in several recent papers
\cite {BW,BW2}.
Because of our exploitation of the underlying group structure,
the formulation we derive is more efficient, in the sense
that it requires integration of fewer coupled equations at a given
order in $1/N$.

A major question which we discuss, but do not fully resolve,
is the propagation of errors induced by truncating the exact
(infinite) hierarchy at a given order in $1/N$.
It is known that the $N\to\infty$ limit is not uniform in time.
For example, in typical large $N$ field theories the characteristic
time scales for scattering or thermalization are known to scale as $N$ to
some positive power.%
\footnote {See, for example, the end of section III of Ref.~\cite {A&Y}.}
For a fixed time interval $t$,
results obtained by integrating evolution equations truncated at,
for example, next-to-leading order, will have only order $1/N^2$ errors.
For sufficiently large $N$, and fixed $t$, including successively higher
orders in the $1/N$ hierarchy will yield more accurate results.
But for fixed $N$ and some given truncation of the $1/N$ hierarchy,
it should be expected that the truncation error will grow with increasing
time and eventually become order unity.
A key question is how this ``breakdown'' time scales with $N$
and the order of the truncation.
One might hope that a next-to-leading order approximation would be
useful [that is, have at most ${\cal O}(1/N)$ global errors] for times
of order $N$, while a next-to-next-to-leading order scheme would
be useful out to times of order $N^2$, {\em etc}.
But it is quite conceivable that errors in an order-$k$ truncation
will grow with time like $(t^\alpha/N)^k$
for some positive $\alpha$,
which would imply that all truncations
break down after a time of order $N^{1/\alpha}$.
This behavior, which we consider likely,
may well depend on the specific theory and choice of
initial state.
Available numerical work, such as \cite {BW,BW2},
sheds little light on this issue.
We discuss several examples where it is possible to argue that
quantum ``decoherence'' produces exactly this type of limit
on the range of validity of large $N$ truncations.

The paper is arranged as follows.
The general framework which allows us to treat
many theories with a classical limit in a uniform fashion is outlined
in section II.
This material is largely taken from Ref.~\cite {Yaffe}.
Section III describes
the particular class of operators we will consider,
and examines the structure of their coherent state equal time correlators.
Section IV presents
the resulting time evolution equations
and discusses error propagation.
These general results are applied to the examples of
point particle quantum mechanics, and a general $N$-component vector model,
in section V.
For point particle quantum mechanics, we argue that the decoherence time
generically scales as $\hbar^{-1/2}$, while for vector models it should
scale as $N^{1/2}$.
A brief concluding discussion follows.

\section{Coherence Group and Coherent States} \label{definitions}

The following slightly abstract framework is applicable to
typical large $N$ limits (including $O(N)$ or $U(N)$ invariant vector models,
matrix models, and non-Abelian gauge theories),
as well as the $\hbar\to0$ limit of ordinary quantum mechanics \cite {Yaffe}.

Consider a quantum theory depending on some
parameter $\chi$ (such as $\hbar$ or $1/N$).
The Hilbert space (which may depend on $\chi$) will be denoted
${\cal H}_\chi$.
The quantum dynamics is governed by a Hamiltonian which we will
write as $(\hbar / \chi) \, \hat H_\chi$.
This rescaling of the Hamiltonian will prove to be convenient,
and makes the Heisenberg equations of motion take the form
\begin {equation}
    {d\over dt} \, \hat A =
    {i\over\chi} \left[ \hat H_\chi , \hat A \right] .
\label {eq:Heis}
\end {equation}

The following assumptions are a set of sufficient conditions
implying that the $\chi\to0$ limit is a classical limit.

Assume there is a Lie group $G$ (called the coherence group) which,
for every value of $\chi$, has a unitary representation on ${\cal H}_\chi$,
$G_\chi = \left\{ D_\chi(u) : u \in G \right\}$.
The states generated by applying elements of the coherence
group to some (normalized) base state $| 0 \rangle_\chi \in {\cal H}_\chi$,
\begin {equation}
    | u \rangle_\chi \equiv D_\chi(u) \, | 0 \rangle_\chi\,, \qquad u \in G \,,
\end {equation}
are called coherent states.
The coherence group acts on these states in a natural way,
$D_\chi(u') \, | u \rangle_\chi = | u' u \rangle_\chi$.

We assume that the coherence group $G_\chi$ acts irreducibly
on the corresponding Hilbert space ${\cal H}_\chi$.
In other words,
no operator (except the identity) commutes with all elements of the
coherence group.
This condition automatically implies that
the set of coherent states form an over-complete basis for the Hilbert space
${\cal H}_\chi$.
It also implies that any operator
acting on ${\cal H}_\chi$ may be
represented as a linear combination of elements of the
coherence group.

For any operator $\hat A$ acting in ${\cal H}_\chi$,
we define its {\em symbol} $A_\chi(u)$ as the set of
coherent state expectation values,
$A_\chi(u) = \langle u| \hat A | u \rangle_\chi$, $u \in G$.
We assume that the only operator whose symbol vanishes identically is
the null operator.
Thus, distinct operators have different symbols, which means
that any operator can, in principle, be completely reconstructed
solely from its diagonal matrix elements in the coherent state basis.

Classical observables will be associated with operators that
remain non-singular as $\chi$ goes to zero,
that is, whose coherent state matrix elements,
$\langle u| \hat A | u' \rangle_\chi / \langle u| u' \rangle_\chi$,
do not blow up as $\chi \rightarrow 0$ for all $u, u' \in G$.
Such operators are called classical.

Two coherent states $| u \rangle$ and $| u' \rangle$ are termed
classically equivalent (we will write $u \sim u'$) if
in the $\chi \rightarrow 0$ limit, one can not
distinguish between them using only classical operators,
{\em i.e.},
$\lim_{\chi \rightarrow 0} A_\chi(u) = \lim_{\chi \rightarrow 0} A_\chi(u')$
for all classical operators $\hat A$.
We assume that the overlap between any two classically inequivalent
coherent states decreases exponentially with $1/\chi$ in the
$\chi \rightarrow 0$ limit.

Under these assumptions,
one may show that the $\chi \rightarrow 0$ limit of this theory
truly is a classical limit \cite {Yaffe}.
The assumptions hold for $O(N)$ or $U(N)$ invariant vector models,
matrix models, and gauge theories \cite {Yaffe}.
The quantum dynamics reduces to classical dynamics on a
phase space $\Gamma$ given by a coadjoint orbit of the coherence group.
Formally, points in
$\Gamma$ correspond to equivalence classes of coherent states,
$\Gamma = \{ [u] : u \in G \}$,
with $[u] = \left\{ u' \in G : u \sim u' \right\}$.
The symplectic structure on the phase space is completely determined
by the Lie algebra structure of the coherence group.
The classical Hamiltonian is just the $\chi\to0$ limit of the coherent
state expectation of the quantum Hamiltonian,
\begin {equation}
    h_{\rm cl}(u) = \lim_{\chi\rightarrow0} \hat H_\chi(u) \,.
    \label{classical hamiltonian}
\end {equation}
To have sensible classical dynamics this limit must exist,
{\em i.e.}, $\hat H_\chi$ must be a classical operator.
(This is why it was convenient the rescale the Hamiltonian by $\hbar/\chi$.)
The classical action is
\begin{equation}
    S_{\rm cl} [u(t)] = \lim_{\chi\to0}
    \int dt \>
    \langle u(t)| \, i \chi \, \partial_t - \hat H_\chi
    |u(t) \rangle_\chi \,.
    \label{coherent state action}
\end{equation}
Both the classical Hamiltonian (\ref {classical hamiltonian})
and the action (\ref {coherent state action})
depend only on the equivalence class of the coherent state $|u\rangle$,%
\footnote
    {%
    For the action,
    this is true up to temporal boundary terms which do not affect the dynamics.
    }
and thus do define sensible dynamics on the classical phase space.

The preceding discussion is just a formalization of the
usual picture of a classical limit.
A quantum mechanical wave packet, with a width of order $\chi^{1/2}$,
behaves classically in the $\chi\to0$ limit, and may be associated
with a point in the classical phase space.
The equations of motion that govern the classical dynamics
are just coherent state expectations
of the original quantum evolution equations.

\section {Coherent State Expectations}
\label{symbols-of-functions}

As noted earlier, the irreducibility of the coherence group implies
that all operators may be (formally) constructed from the generators
of the coherence group.
Consequently,
for characterizing the structure, and time evolution, of any state,
one may focus attention on equal-time expectation values
of products of coherence group generators.

Let $g$ denote the Lie algebra of the coherence group $G$.
Let $\{ e_i \}$ be a basis of $g$.
The commutator of basis elements defines
the structure constants, 
$\left[ e_i , e_j \right] = i f_{ij}^k \, e_k$.
The generators $e_i$ themselves are not classical operators,
but rather are $1/\chi$ times classical operators.
For convenience, let $\hat x_i$ denote the rescaled generator
which is a classical operator,
$\hat x_i \equiv \chi e_i$.

Consider the coherent state expectation value of the monomial
$\hat x_{i_1} \hat x_{i_2} \cdots \hat x_{i_k}$.
We would like to find an expansion of this expectation value
in powers of $\chi$.
A convenient representation for our purposes involves subtracted expectations%
\footnote
    {
    To simplify notation, we will omit the superscript ``$(k)$''
    when this can cause no confusion; for example, we will
    write $g_{ij}$ for $g^{(2)}_{ij}$, {\em etc.}
    The same remark applies to the connected expectations discussed below.
    }
\begin{eqnarray}
    g^{(k)}_{i_1 i_2 \cdots i_k} &\equiv&
    \langle (\hat x_{i_1}- x_{i_1})\cdots (\hat x_{i_k} - x_{i_k})
    \rangle \,,
    \label{subtracted-expectations-definition}
\end{eqnarray}
where $\langle \cdots  \rangle$ denotes an expectation in
some coherent state, and
$x_i \equiv \langle \hat x_i \rangle$ are the expectations
of the rescaled generators $\hat x_i$.
Subtracted and un-subtracted expectations are related by
\begin{eqnarray}
    \langle \hat x_{i_1} \cdots  \hat x_{i_k} \rangle
    &=& x_{i_1} \cdots  x_{i_k}
    \times \left\{ 1
    + \sum_{(l_1, l_2)}
    {g^{(2)}_{l_1 l_2} \over x_{l_1} x_{l_2} }
    + \sum_{(l_1, l_2, l_3)}
    {g^{(3)}_{l_1 l_2 l_3} \over x_{l_1} x_{l_2} x_{l_3} }
    + \cdots  +
    {g^{(k)}_{i_1 \cdots  i_k} \over x_{i_1} \cdots  x_{i_k} }
    \right\} ,
    \label{generators-product}
\end{eqnarray}
where the $n$-tuples $(l_1, l_2, ... , l_n)$ are ordered subsets of
$\{i_1 , ... , i_k\}$.
(There is no $g^{(1)}$ term since
$g^{(1)}_i \equiv \langle \hat x_i - x_i \rangle = 0$.)

\begin{figure}[t!]
\vskip -0.1in
(a)
\vskip -0.2in
\hskip 0.3in \epsfig{height=0.85in, file=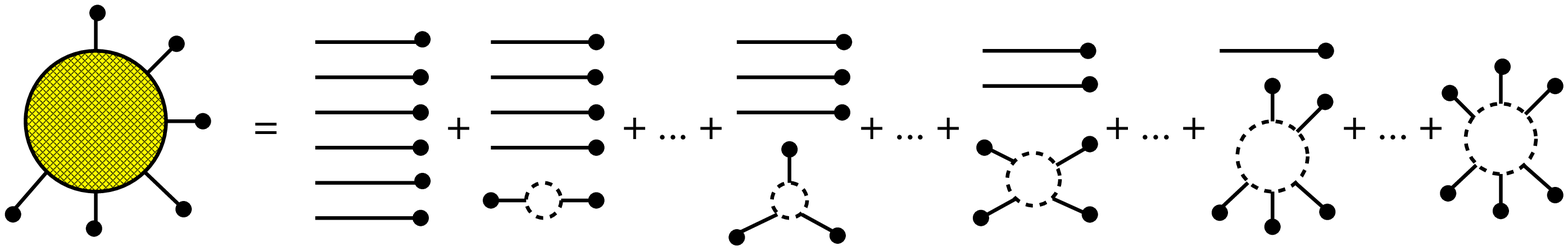}
\vskip 0.05in
(b)
\vskip -0.2in
\hskip 0.3in \epsfig{height=1.8in, file=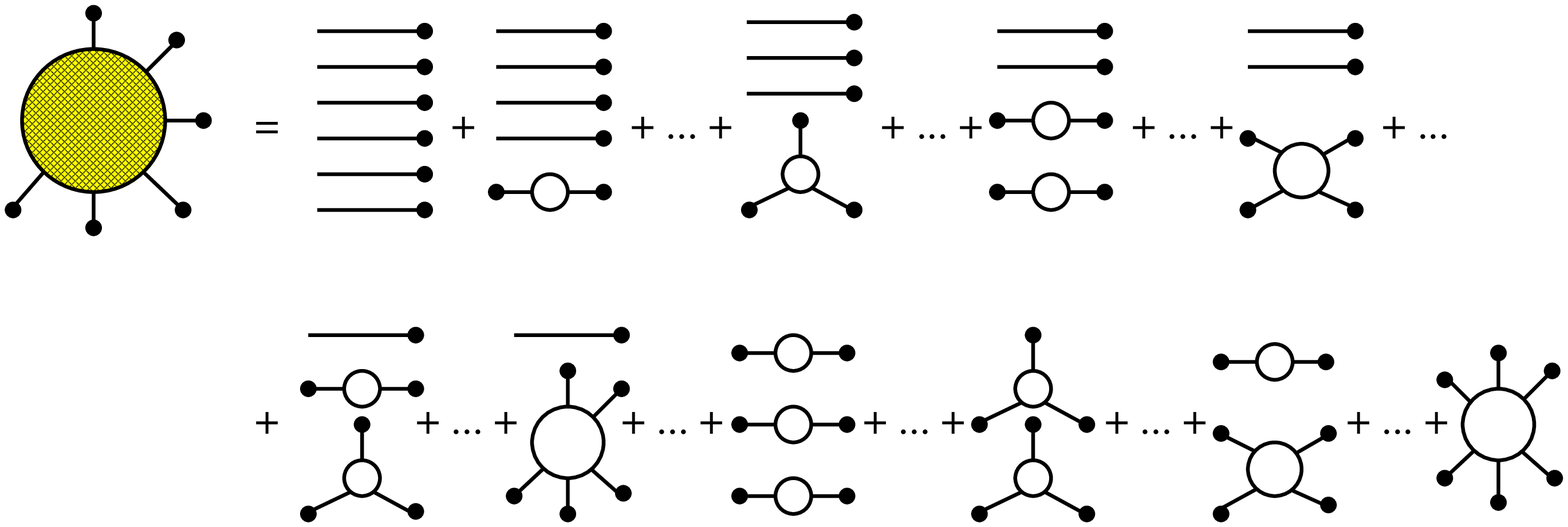}
\vskip 0.1in
\caption{%
Expansion of the expectation a product of generators in terms of
(a) subtracted, and (b) connected expectations.
The shaded bubbles on the left denote full expectations of
the product of generators.
Each 
line with a dot on the end represents the
expectation of a single generator.
In (a), dashed-line bubbles correspond to
subtracted $g^{(k)}$ expectations of strings of generators,
while in (b) solid-line bubbles represent
connected $s^{(k)}$ expectations.
Ellipses ($\cdots$) denote ordered
permutations of the preceding diagram.
\label {fig:sub-vs-con}
}
\end {figure}

Alternatively, one may expand in terms of connected expectations,
\begin {equation}
    s^{(k)}_{i_1 \cdots i_k} \equiv
    \langle \hat x_{i_1} \cdots \hat x_{i_k} \rangle^{\rm conn} \,.
\end {equation}
The difference,
illustrated graphically in Fig.~\ref {fig:sub-vs-con},
is that
expansions in terms of connected expectations involve products of all
possible `contractions', while the terms in the expansion in subtracted
expectations have only one string of generators `contracted'.
The difference between subtracted and connected expectations
first arises with four generators.
Explicitly,
\begin {mathletters}
\begin{eqnarray}
\langle \hat x_i \hat x_j \hat x_k \hat x_l \rangle &=&
x_i x_j x_k x_l +
x_i x_j \, g^{(2)}_{k l} +
x_i x_k \, g^{(2)}_{j l} +
x_i x_l \, g^{(2)}_{j k} +
x_k x_l \, g^{(2)}_{i j} +
x_j x_l \, g^{(2)}_{i k} +
x_j x_k \, g^{(2)}_{i l} \nonumber\\&&\quad {} +
x_i \, g^{(3)}_{j k l} +
x_j \, g^{(3)}_{i k l} +
x_k \, g^{(3)}_{i j l} +
x_l \, g^{(3)}_{i j k} +
g^{(4)}_{i j k l} \\&=&
x_i x_j x_k x_l +
x_i x_j \, s^{(2)}_{k l} +
x_i x_k \, s^{(2)}_{j l} +
x_i x_l \, s^{(2)}_{j k} +
x_k x_l \, s^{(2)}_{i j} +
x_j x_l \, s^{(2)}_{i k} +
x_j x_k \, s^{(2)}_{i l} \nonumber\\&&\quad {} +
x_i \, s^{(3)}_{j k l} +
x_j \, s^{(3)}_{i k l} +
x_k \, s^{(3)}_{i j l} +
x_l \, s^{(3)}_{i j k} +
s^{(2)}_{i j} s^{(2)}_{k l} +
s^{(2)}_{i k} s^{(2)}_{j l} +
s^{(2)}_{i l} s^{(2)}_{j k} +
s^{(4)}_{i j k l} \,.
\end{eqnarray}
\end {mathletters}

The coherent state overlap $\langle u|u'\rangle_\chi$
is the generating functional
for expectations of products of generators, since variations of
the coherent state $u'$ can bring down any desired generator of
the Lie algebra, $\delta_i \, |u'\rangle = e_i |u'\rangle$.
The logarithm of this overlap is therefore the generating functional
for connected expectations.
By assumption, $\ln \langle u|u'\rangle_\chi$ is ${\cal O}(1/\chi)$ as $\chi\to 0$.
This immediately implies that the
$k$-th order connected expectation $s^{(k)}$ is ${\cal O}(\chi^{k-1})$.
Note also that the commutator of functional derivatives is the
functional derivative in the direction of the commutator,
\begin {equation}
\langle \cdots \hat x_i \hat x_j \cdots \rangle^{\rm conn} -
\langle \cdots \hat x_j \hat x_i \cdots \rangle^{\rm conn}
= \langle \cdots \left[\hat x_i , \hat x_j \right] \cdots \rangle^{\rm conn}
= i \chi \, f_{i j}^m \, \langle \cdots \hat x_m \cdots \rangle^{\rm conn} ,
\end {equation}
or
$s^{(k)}_{\cdots i j \cdots} - s^{(k)}_{\cdots j i \cdots} =
i \chi \, f_{i j}^m \, s^{(k-1)}_{\cdots m \cdots }$.

By considering which connected expectations contribute to  $g^{(k)}$,
one may easily see%
\footnote{
As $g^{(k)} = \sum s^{(m_1)} s^{(m_2)} \cdots s^{(m_n)}$ with
$m_1 + \cdots + m_n = k$ and $m_i > 1$ for all $i$, and
$s^{(m_i)} = {\cal O}(\chi^{m_i - 1})$, $g^{(k)} =
{\cal O}(\chi^{\min\{ \sum (m_i-1) \}})
= {\cal O}(\chi^{\min\{ k - n \}})$.
The largest number of connected diagrams $n$ occurs when
all $s^{(m_i)}$ are $s^{(2)}$ (except for one, if $k$ is odd,
which is $s^{(3)}$).
}
that subtracted expectations fall off roughly half as fast as the
connected ones, $g^{(2k)} \sim g^{(2k-1)}= {\cal O}(\chi^k)$.
Because $g^{(k+2)} / g^{(k)} = {\cal O}(\chi)$, expansion
(\ref{generators-product}) is a power series in $\chi$, the
parameter measuring how close the system is to being classical.
Of course, subtracted expectations may always be rewritten in terms
of connected expectations (and vice-versa).
Ultimately, equations for connected%
\footnote
    {%
    Or perhaps one-particle irreducible.
    }
expectations will be most useful.
Nevertheless, using subtracted expectations as an intermediate
representation is helpful because of the simple form of
expansions in terms of subtracted expectations, as shown by
Eq.~(\ref {generators-product}) and Eq.~(\ref{eq:V-sub})
below.
For later use, note that
\begin {equation}
    g^{(k)}_{ \cdots b i j c \cdots } - g^{(k)}_{ \cdots b j i c \cdots }
    = i \chi \, f^m_{i j}
	\left( x_m \, g^{(k-2)}_{ \cdots b c \cdots } +
	g^{(k-1)}_{ \cdots b m c \cdots } \right) .
\label {eq:g-red}
\end {equation}

Now consider an operator $V$ that can (at least formally)
be expanded in powers of generators,
$
    V = \sum_k \sum_{\{i_1 \cdots i_k\}}
    \alpha_{i_1 \cdots i_k} \, \hat x_{i_1} \cdots \hat x_{i_k}
$,
for some set of coefficients $\{ \alpha_{i_1 \cdots i_k}\}$.
Operators of this form are well behaved for
$\chi \rightarrow0$, and so are good classical operators.
Using (\ref{generators-product}),
\begin{eqnarray}
    \langle V \rangle &=&
    \sum_k \!
    \sum_{\{i_1 \cdots i_k\}}
    \alpha_{i_1 \cdots i_k} \,
    x_{i_1} \cdots x_{i_k}
    \left\{ 1
    +
    \sum_{(l_1, l_2)}
    {g^{(2)}_{l_1 l_2} \over x_{l_1} x_{l_2}}
    + \sum_{(l_1, l_2, l_3)}
    {g^{(3)}_{l_1 l_2 l_3} \over x_{l_1} x_{l_2} x_{l_3}}
    + \cdots +
    {g^{(k)}_{i_1 \cdots i_k} \over x_{l_1} \cdots x_{l_k}}
    \right\}.
    \label {eq:V-sub}
\end{eqnarray}
This can be packaged
in an even more concise form,
\begin{equation}
    \langle V
    \rangle = \bar V
    +
    g^{(2)}_{l_1 l_2} \,
    V^{(l_1 l_2)}
    +
    g^{(3)}_{l_1 l_2 l_3} \,
    V^{(l_1 l_k l_3)}
    +
    g^{(4)}_{l_1 l_2 l_3 l_4} \,
    V^{(l_1 l_k l_3 l_4)}
    +
    {\cal O}(\chi^3)
    \label{power-series-expansion-symbol}
\end{equation}
where summation on repeated indices is implied,
and
$
    \bar V \equiv \sum_k \sum_{\{i_1 \cdots i_k\}}
    \alpha_{i_1 \cdots i_k} \, x_{i_1} \cdots x_{i_k}
$
is the number obtained by replacing each generator in $V$
by its coherent state expectation.
Here we have introduced
``ordered derivatives''
$f^{(i j \cdots )} \equiv {\delta f \over \delta (x_i x_j \cdots )}$
defined by
\begin {mathletters}
\begin{eqnarray}
    f^{(l)} &=&
    {\partial \bar f \over \partial x_{l}} \,,
\\
    (f g)^{(l_1 \cdots l_k)} &=&
    \sum_{i=0}^{k} f^{(l_1 \cdots l_i)} g^{(l_{i+1} \cdots l_k)} \,.
    %
\end{eqnarray}%
\label{operator-derivatives}%
\end {mathletters}
When acting on a string of generators, ordered derivatives
produce a sum of products of expectations of the generators
which remain after deleting
the indicated generators, provided these appear
(not necessarily contiguously)
somewhere within the string in the order specified by the derivative.
For example,%
\footnote
    {
    If $\hat x_{i_1}, \cdots , \hat x_{i_n}$ commute, then
    $
	{\delta f \over \delta (x_{i_1} \cdots x_{i_n})}
	=
	{1 \over n!} \,
	{\partial^n \bar f \over \partial x_{i_1} \cdots \partial x_{i_n}}
    $.
    In this case, the ordering does not matter, and the $n!$ is needed
    to make up for over-counting.
    }
${\delta \over \delta (x p)} \, \hat x^2 \hat p \hat x = 2 x^2$,
and
$
    {\delta \over \delta (x p)} \, \hat x \hat p \hat x^2 \hat p
    =
    x^2 p + 2 xpx + p x^2
    =
    4 x^2 p
$.

In the $\chi\to0$ limit,
coherent state expectations of the (rescaled) generators
$\hat x_i$ turn into coordinates $x_i$ on the classical
phase space and (classical) operators acting on ${\cal H}_\chi$
become functions on phase space,
$\langle V\rangle = \bar V + O(\chi)$.
For finite $\chi$,
the successive terms in (\ref{power-series-expansion-symbol})
precisely characterize the corrections to this classical limit.

\section{Time Evolution} \label{time-evolut}

Since operators are completely determined by their symbols,
to study the time dependence of any observable $\hat A$
it is sufficient to take
the coherent state expectation value of its Heisenberg equation of motion
(\ref {eq:Heis}),
\begin {equation}
    {d\over dt} \, \langle \hat A\rangle =
    {i\over\chi}
    \left\langle  \left[ \hat H_\chi , \hat A \right] \right\rangle .
\end {equation}
In other words, we assume that the initial state is precisely some
coherent state $|u\rangle$, and wish to determine the subsequent time evolution.
To do so,
we will first find an expansion, in powers of $\chi$,
for the expectation of the commutator of classical operators.

\subsection{Symbols of Commutators}

Consider classical operators $A$ and $B$ which
(as in Section \ref{symbols-of-functions})
may be written as power
series in the generators,
$A = \sum \alpha_{i_1 \cdots i_m} \, \hat x_{i_1} \cdots \hat x_{i_m}$,
$B = \sum \beta_{j_1 \cdots j_n} \, \hat x_{j_1} \cdots \hat x_{j_n}$.
Their product is given by
$A B = \sum \alpha_{i_1 \cdots i_m} \beta_{j_1 \cdots j_n}
\hat x_{i_1} \cdots \hat x_{i_m} \hat x_{j_1} \cdots \hat x_{j_n}$.
Using our previous result (\ref{power-series-expansion-symbol}),
we find
\begin {mathletters}
\begin{eqnarray}
    \langle A B \rangle &=&
    \bar A \bar B
    + g^{(2)}_{l_1 l_2} \, (A B)^{(l_1 l_2)}
    + g^{(3)}_{l_1 l_2 l_3} \, (A B)^{(l_1 l_2 l_3)}
    + g^{(4)}_{l_1 l_2 l_3 l_4} \, (A B)^{(l_1 l_2 l_3 l_4)}
    + {\cal O}(\chi^3)
\\ &=&
    \bar A \bar B
    + s^{(2)}_{l_1 l_2} \, (A B)^{(l_1 l_2)}
    + s^{(3)}_{l_1 l_2 l_3} \, (A B)^{(l_1 l_2 l_3)}
\nonumber\\&& \qquad\qquad\qquad{}
    + \left(
	s^{(2)}_{l_1 l_2} \, s^{(2)}_{l_3 l_4} +
	s^{(2)}_{l_1 l_3} \, s^{(2)}_{l_2 l_4} +
	s^{(2)}_{l_1 l_4} \, s^{(2)}_{l_2 l_3}
    \right)
    (A B)^{(l_1 l_2 l_3 l_4)}
    + {\cal O}(\chi^3)
\end{eqnarray}
\label{product-of-operators}%
\end{mathletters}
where now $(l_1, l_2, \cdots )$ denote ordered subsets of
$\{i_1, \cdots , i_m, j_1, \cdots , i_n\}$.
We see from (\ref{product-of-operators})
that, to leading order, products of classical operators factorize,
$\langle A B \rangle=\langle A \rangle \langle B \rangle +{\cal O}(\chi)$.

Using the expansion (\ref{product-of-operators}),
and the reduction formulas for
operator derivatives (\ref{operator-derivatives}),
one can evaluate the commutator.
A generic term in the result will be
\begin{eqnarray}
\label{generic-term-in-gk}
    g^{(k)}_{l_1 l_2 \cdots l_k}
    \left[
    (A B)^{(l_1 l_2 \cdots l_k)}
    -
    (B A)^{(l_1 l_2 \cdots l_k)}
    \right]
    \hspace{-11em}
&& \nonumber \\
&=& \phantom{-\;}
    g^{(k)}_{l_1 \cdots l_k}
    \left(
    A^{(l_1 \cdots l_k)} \bar B
    + \cdots
    + A^{(l_1 l_2)} B^{(l_3 \cdots l_k)}
    + A^{(l_1)} B^{(l_2 \cdots l_k)}
    + \bar A B^{(l_1 \cdots l_k)}
    \right)
    \nonumber \\ && {} -
    g^{(k)}_{l_1 \cdots l_k}
    \left(
    B^{(l_1 \cdots l_k)} \bar A
    + \cdots
    + B^{(l_1 l_2)} A^{(l_3 \cdots l_k)}
    + B^{(l_1)} A^{(l_2 \cdots l_k)}
    + \bar B A^{(l_1 \cdots l_k)}
    \right)
    \nonumber \\ &=& \phantom{+}
    \left(
    g^{(k)}_{l_1 l_2 l_3 l_4 \cdots l_k} - g^{(k)}_{l_2 l_3 l_4 \cdots l_k l_1}
    \right)
    \left(
    A^{(l_1)} B^{(l_2 l_3 l_4 \cdots l_k)} -
    B^{(l_1)} A^{(l_2 l_3 l_4 \cdots l_k)}
    \right)
    \nonumber \\ && +
    \left(
    g^{(k)}_{l_1 l_2 l_3 l_4 \cdots l_k} - g^{(k)}_{l_3 l_4 \cdots l_k l_1 l_2}
    \right)
    \left(
    A^{(l_1 l_2)} B^{(l_3 l_4 \cdots l_k)} -
    B^{(l_1 l_2)} A^{(l_3 l_4 \cdots l_k)}
    \right)
    \nonumber \\ && +
    \left(
    g^{(k)}_{l_1 l_2 l_3 l_4 \cdots l_k} - g^{(k)}_{l_4 \cdots l_k l_1 l_2 l_3}
    \right)
    \left(
    A^{(l_1 l_2 l_3)} B^{(l_4 \cdots l_k)} -
    B^{(l_1 l_2 l_3)} A^{(l_4 \cdots l_k)}
    \right)
    + \cdots ~.
\end{eqnarray}
The last term in the sum (\ref{generic-term-in-gk}) is either
$
    \left(
    g^{(2j)}_{l_1 \cdots l_j l_{j+1} \cdots l_{2j}} -
    g^{(2j)}_{l_{j+1} \cdots l_{2j} l_1 \cdots l_j}
    \right)
    \left(
    A^{(l_1 \cdots l_j)} B^{(l_{j+1} \cdots l_{2j})}
    \right)
$
or
$
    \left(
    g^{(2j+1)}_{l_1 \cdots l_j l_{j+1} \cdots l_{2j+1}} -
    g^{(2j+1)}_{l_{j+1} \cdots l_{2j+1} l_1 \cdots l_j}
    \right)
$$
    \left(
    A^{(l_1 \cdots l_j)} B^{(l_{j+1} \cdots l_{2j+1})} -
    B^{(l_1 \cdots l_j)} A^{(l_{j+1} \cdots l_{2j+1})}
    \right) ,
$
depending on whether $k$ is even or odd.
Using (\ref {eq:g-red}) to reduce
the differences $(g^{(k)}_{\cdots} - g^{(k)}_{\cdots})$
yields the final form for the expectation of the commutator of
classical operators.
The leading term is precisely the Poisson bracket on the
classical phase space,
while subsequent terms involve successively higher expectations
$g^{(k)}$.
Displaying subleading ${\cal O}(\chi)$ and ${\cal O}(\chi^2)$
terms explicitly, one finds
\begin{eqnarray}
\label{corrected-commutator}
\langle {1 \over i \chi} \left[ A , B \right] \rangle
\hspace{-4ex} && \nonumber \\
&& \matrix{
=
\left[ f^m_{l_1 l_2} x_m \right] \left( A^{(l_1)} B^{(l_2)} \right)
\hfill
}
\hspace{52.73ex} \Bigr\} \>
{\cal O}(\chi^0) \nonumber\\
&& \phantom{\small|} \nonumber\\
&&\left. \matrix{
+
\left[
f^m_{l_1 l_2} \, g^{}_{m l_3} + f^m_{l_1 l_3}\, g^{}_{l_2 m}
\right]
\left( A^{(l_1)} B^{(l_2 l_3)} - B^{(l_1)} A^{(l_2 l_3)} \right)
\hfill \cr
+
\left[
x_m \left(
f^m_{l_1 l_2}\, g^{}_{l_3 l_4} +
f^m_{l_1 l_3}\, g^{}_{l_2 l_4} +
f^m_{l_1 l_4}\, g^{}_{l_2 l_3}
\right)
\right]
\left( A^{(l_1)} B^{(l_2 l_3 l_4)} - B^{(l_1)} A^{(l_2 l_3 l_4)} \right)
\hfill \cr
+
\left[
x_m \left(
f^m_{l_1 l_3}\, g^{}_{l_2 l_4} +
f^m_{l_1 l_4}\, g^{}_{l_3 l_2} +
f^m_{l_2 l_3}\, g^{}_{l_1 l_4} +
f^m_{l_2 l_4}\, g^{}_{l_3 l_1}
\right)
\right]
\left( A^{(l_1 l_2)} B^{(l_3 l_4)} \right)
\hfill
}
\hspace{6.78ex} \right\} {\cal O}(\chi^1) \nonumber\\
&& \phantom{\Big|} \nonumber\\
&& \left. \matrix{
+
\left[
f^m_{l_1 l_2}\, g^{}_{m l_3 l_4} +
f^m_{l_1 l_3}\, g^{}_{l_2 m l_4} +
f^m_{l_1 l_4}\, g^{}_{l_2 l_3 m}
\right]
\left( A^{(l_1)} B^{(l_2 l_3 l_4)} - B^{(l_1)} A^{(l_2 l_3 l_4)} \right)
\hfill \cr
+
\left[
f^m_{l_1 l_3}\, g^{}_{m l_2 l_4} +
f^m_{l_1 l_4}\, g^{}_{l_3 m l_2} +
f^m_{l_2 l_3}\, g^{}_{l_1 m l_4} +
f^m_{l_2 l_4}\, g^{}_{l_3 l_1 m}
\right]
\left( A^{(l_1 l_2)} B^{(l_3 l_4)} \right)
\hfill \cr
+
\left[
x_m \left(
f^m_{l_1 l_2}\, g^{}_{l_3 l_4 l_5} +
f^m_{l_1 l_3}\, g^{}_{l_2 l_4 l_5} +
f^m_{l_1 l_4}\, g^{}_{l_2 l_3 l_5} +
f^m_{l_1 l_5}\, g^{}_{l_2 l_3 l_4}
\right)
\right.
\hfill \cr
\left. \hspace{3ex}
+
f^m_{l_1 l_2}\, g^{}_{m l_3 l_4 l_5} +
f^m_{l_1 l_3}\, g^{}_{l_2 m l_4 l_5} +
f^m_{l_1 l_4}\, g^{}_{l_2 l_3 m l_5} +
f^m_{l_1 l_5}\, g^{}_{l_2 l_3 l_4 m}
\right]
\hfill \cr
\hspace{15em} \times
\left(
A^{(l_1)}
B^{(l_2 l_3 l_4 l_5)}
-
B^{(l_1)}
A^{(l_2 l_3 l_4 l_5)}
\right)
\hfill \cr
+
\left[
x_m \left(
f^m_{l_1 l_3}\, g^{}_{l_4 l_5 l_2} +
f^m_{l_1 l_4}\, g^{}_{l_3 l_5 l_2} +
f^m_{l_1 l_5}\, g^{}_{l_3 l_4 l_2}
+
f^m_{l_2 l_3}\, g^{}_{l_1 l_4 l_5} +
f^m_{l_2 l_4}\, g^{}_{l_1 l_3 l_5}
\right.
\right. \hfill \cr
\left. \hspace{3ex}
\left.
+ f^m_{l_2 l_5}\, g^{}_{l_1 l_3 l_4}
\right)
+
f^m_{l_1 l_3}\, g^{}_{m l_4 l_5 l_2} +
f^m_{l_1 l_4}\, g^{}_{l_3 m l_5 l_2} +
f^m_{l_1 l_5}\, g^{}_{l_3 l_4 m l_2} +
f^m_{l_2 l_3}\, g^{}_{l_1 m l_4 l_5}
\right. \hfill \cr
\left. \hspace{3ex}
+ f^m_{l_2 l_4}\, g^{}_{l_1 l_3 m l_5}
+ f^m_{l_2 l_5}\, g^{}_{l_1 l_3 l_4 m}
\right]
\left(
A^{(l_1 l_2)}
B^{(l_3 l_4 l_5)}
-
B^{(l_1 l_2)}
A^{(l_3 l_4 l_5)}
\right)
\hfill \cr
+
\left[
x_m \left(
f^m_{l_1 l_2}\, g^{}_{l_3 l_4 l_5 l_6} +
f^m_{l_1 l_3}\, g^{}_{l_2 l_4 l_5 l_6} +
f^m_{l_1 l_4}\, g^{}_{l_2 l_3 l_5 l_6} +
f^m_{l_1 l_5}\, g^{}_{l_2 l_3 l_4 l_6} +
f^m_{l_1 l_6}\, g^{}_{l_2 l_3 l_4 l_5}
\right)
\right]
\hfill \cr
\hspace{15em} \times
\left(
A^{(l_1)}
B^{(l_2 l_3 l_4 l_5 l_6)}
-
B^{(l_1)}
A^{(l_2 l_3 l_4 l_5 l_6)}
\right)
\hfill \cr
+
\left[
x_m \left(
f^m_{l_1 l_3}\, g^{}_{l_4 l_5 l_6 l_2} +
f^m_{l_1 l_4}\, g^{}_{l_3 l_5 l_6 l_2} +
f^m_{l_1 l_5}\, g^{}_{l_3 l_4 l_6 l_2} +
f^m_{l_1 l_6}\, g^{}_{l_3 l_4 l_5 l_2}
\right. \right. \hfill \cr \left. \left. \hspace{3ex} +
f^m_{l_2 l_3}\, g^{}_{l_1 l_4 l_5 l_6} +
f^m_{l_2 l_4}\, g^{}_{l_1 l_3 l_5 l_6} +
f^m_{l_2 l_5}\, g^{}_{l_1 l_3 l_4 l_6} +
f^m_{l_2 l_6}\, g^{}_{l_1 l_3 l_4 l_5}
\right)
\right]
\hfill \cr \hspace{15em} \times
\left(
A^{(l_1 l_2)}
B^{(l_3 l_4 l_5 l_6)}
-
B^{(l_1 l_2)}
A^{(l_3 l_4 l_5 l_6)}
\right)
\hfill \cr
+
\left[
x_m \left(
f^m_{l_1 l_4}\, g^{}_{l_5 l_6 l_2 l_3} +
f^m_{l_1 l_5}\, g^{}_{l_4 l_6 l_2 l_3} +
f^m_{l_1 l_6}\, g^{}_{l_4 l_5 l_2 l_3} +
f^m_{l_2 l_4}\, g^{}_{l_1 l_5 l_6 l_3} +
f^m_{l_2 l_5}\, g^{}_{l_1 l_4 l_6 l_3}
\right.\right. \hfill \cr \left.\left. \hspace{3ex} +
f^m_{l_2 l_6}\, g^{}_{l_1 l_4 l_5 l_3} +
f^m_{l_3 l_4}\, g^{}_{l_1 l_2 l_5 l_6} +
f^m_{l_3 l_5}\, g^{}_{l_1 l_2 l_4 l_6} +
f^m_{l_3 l_6}\, g^{}_{l_1 l_2 l_4 l_5}
\right)
\right]
\left( A^{(l_1 l_2 l_3)} B^{(l_4 l_5 l_6)} \right)
\hfill
}
\hfill \right\} {\cal O}(\chi^2)
\nonumber\\ &&
\nonumber\\
&& \left. \matrix{ {}
+
{\cal O}(\chi^3) \,.} \right.
\end{eqnarray}

\subsection{Equations of Motion} \label{e-o-m}
\vspace*{-8pt}

To determine the evolution to order ${\cal O}(\chi^3)$,
we need the time derivatives of
$x_i(t)$, $g_{ij}(t)$, and $g_{ijk}(t)$.
Take the commutator of products of generators with the
Hamiltonian and subtract the disconnected parts to find:
\begin{eqnarray}
\label{x-evolution}
    \hspace{-5ex} {d\over dt} \, x_{i} &=&
    \left[
    f^a_{i j} \, x_a
    \right]
    H^{(j)}
    +
    \left[
    f^a_{i j} \, g_{a k} + f^a_{i k} \, g_{j a}
    \right]
    H^{(j k)}
    \nonumber\\&+&
    \left[
    x_a \left(
    f^a_{i j} \, g_{k l} +
    f^a_{i k} \, g_{j l} +
    f^a_{i l} \, g_{j k}
    \right)
    +
    f^a_{i j} \, g_{a k l} +
    f^a_{i k} \, g_{j a l} +
    f^a_{i l} \, g_{j k a}
    \right]
    H^{(j k l)}
    \nonumber \\&+&
    \left[
    x_a \left(
    f^a_{i j} \, g_{k l m} +
    f^a_{i k} \, g_{j l m} +
    f^a_{i l} \, g_{j k m} +
    f^a_{i m} \, g_{j k l}
    \right)
    \right. \nonumber \\ && \left. \hspace{3ex}
    +
    f^a_{i j}\, g_{a k l m} +
    f^a_{i k}\, g_{j a l m} +
    f^a_{i l}\, g_{j k a m} +
    f^a_{i m}\, g_{j k l a}
    \right]
    H^{(j k l m)}
    \nonumber \\&+&
    \left[
    x_a \left(
    f^a_{i j}\, g_{k l m n}
    +
    f^a_{i k}\, g_{j l m n} +
    f^a_{i l}\, g_{j k m n} +
    f^a_{i m}\, g_{j k l n} +
    f^a_{i n}\, g_{j k l m}
    \right)
    \right]
    H^{(j k l m n)}
    \nonumber\\&+&
    {\cal O}(\chi^3)
    \\
    \label{xx-evolution}
    \hspace{-5ex} {d\over dt} \, g_{i j} &=&
    \left[
    f^a_{i k}\, g_{a j} + f^a_{j k}\, g_{i a}
    \right]
    H^{(k)}
    \nonumber\\&+&
    \left[
    x_a \left(
    f^a_{i k}\, g_{j l} +
    f^a_{i l}\, g_{k j} +
    f^a_{j k}\, g_{i l} +
    f^a_{j l}\, g_{k i}
    \right)
    +
    f^a_{i k}\, g_{a j l} +
    f^a_{i l}\, g_{k a j} +
    f^a_{j k}\, g_{i a l} +
    f^a_{j l}\, g_{k i a}
    \right]
    H^{(k l)}
    \nonumber \\&+&
    \left[
    x_a \left(
    f^a_{i k}\, g_{l m j} +
    f^a_{i l}\, g_{k m j} +
    f^a_{i m}\, g_{k l j}
    +
    f^a_{j k}\, g_{i l m} +
    f^a_{j l}\, g_{i k m} +
    f^a_{j m}\, g_{i k l}
    \right)
    \right. \nonumber \\ && \left. \hspace{3ex}
    +
    f^a_{i k}\, g_{a l m j} +
    f^a_{i l}\, g_{k a m j}
    +
    f^a_{i m}\, g_{k l a j}
    +
    f^a_{j k}\, g_{i a l m} +
    f^a_{j l}\, g_{i k a m} +
    f^a_{j m}\, g_{i k l a}
    \right]
    H^{(k l m)}
    \nonumber \\&+&
    \left[
    x_a \left(
    f^a_{i k}\, g_{l m n j} +
    f^a_{i l}\, g_{k m n j} +
    f^a_{i m}\, g_{k l n j} +
    f^a_{i n}\, g_{k l m j}
    \right. \right. \nonumber \\&& \left. \left. \hspace{3ex}
    +
    f^a_{j k}\, g_{i l m n} +
    f^a_{j l}\, g_{i k m n} +
    f^a_{j m}\, g_{i k l n} +
    f^a_{j n}\, g_{i k l m}
    \right)
    \right]
    H^{(k l m n)}
    \nonumber\\&+&
    {\cal O}(\chi^3)
    \\
    \label{xxx-evolution}
    \hspace{-5ex} {d\over d t} \, g_{ijk} &=&
    \left[
    f^a_{i l}\, g_{a j k} +
    f^a_{j l}\, g_{i a k} +
    f^a_{k l}\, g_{i j a}
    \right]
    H^{(l)}
    \nonumber \\&+&
    \left[
    x_a \left(
    f^a_{i l}\, g_{j k m} +
    f^a_{j l}\, g_{i k m} +
    f^a_{k l}\, g_{i j m}
    +
    f^a_{i m}\, g_{l j k} +
    f^a_{j m}\, g_{l i k} +
    f^a_{k m}\, g_{l i j}
    \right)
    \right]
    H^{(l m)}
    \nonumber \\&+&
    \left[
    f^a_{i l} (
    g_{a j}\, g_{k m} +
    g_{a k}\, g_{j m} )
    +
    f^a_{j l} (
    g_{i a}\, g_{k m} +
    g_{i m}\, g_{a k} )
    +
    f^a_{k l} (
    g_{i a}\,  g_{j m} +
    g_{i m}\,  g_{a j} )
    \right. \nonumber\\ && \left. \hspace{3ex}
    +
    f^a_{i m} (
    g_{l j}\, g_{a k} +
    g_{l k}\, g_{a j} )
    +
    f^a_{j m} (
    g_{l i}\, g_{a k} +
    g_{l a}\, g_{k i} )
    +
    f^a_{k m} (
    g_{l i}\, g_{j a} +
    g_{l j}\, g_{i a} )
    \right]
    H^{(l m)}
    \nonumber \\&+&
    \left[
    x_a \left(
    f^a_{i l} (
    g_{m j}\, g_{ n k} +
    g_{m k}\, g_{ n j} )
    +
    f^a_{i m} (
    g_{l j}\, g_{ n k} +
    g_{l k}\, g_{ n j} )
    +
    f^a_{i n} (
    g_{l j}\, g_{ m k} +
    g_{l k}\, g_{ m j} )
    \right.\right. \nonumber \\&& \left.\left. \hspace{3ex}
    +
    f^a_{j l} (
    g_{i m}\, g_{ n k} +
    g_{i n}\, g_{ m k} )
    +
    f^a_{j m} (
    g_{i l}\, g_{ n k} +
    g_{i n}\, g_{ l k} )
    +
    f^a_{j n} \left(
    g_{i l}\, g_{ m k} +
    g_{i m}\, g_{ l k} \right)
    \right.\right. \nonumber \\&& \left.\left. \hspace{3ex}
    +
    f^a_{k l} \left(
    g_{i m}\, g_{j n} +
    g_{i n}\, g_{j m} \right)
    +
    f^a_{k m} \left(
    g_{i l}\, g_{j n} +
    g_{i n}\, g_{j l} \right)
    +
    f^a_{k n} \left(
    g_{i l}\, g_{j m} +
    g_{i m}\, g_{j l} \right)
    \right)
    \right]
    H^{(l m n)}
    \nonumber\\&+&
    {\cal O}(\chi^3) .
\end{eqnarray}
Recall that, through third order, there is no difference
between the subtracted and connected correlators.
Only the disconnected parts of the fourth order correlators
appearing in Eq's.~(\ref {x-evolution}) and (\ref {xx-evolution})
are needed, since
$
    g_{i j k l} = g_{i j} g_{k l} + g_{i k} g_{j l} + g_{i l} g_{j k}
    + {\cal O}(\chi^3)
$.
If equations only accurate to ${\cal O}(\chi^2)$ are desired,
then all terms in Eq's.~(\ref {x-evolution})--(\ref {xxx-evolution})
involving third (or higher) order correlators,
as well as products of second order correlators, may be dropped.%
\footnote
    {
    The resulting next-to-leading order equations are simply
    \begin {mathletters}
    \label {eq:NLO}
    \begin {eqnarray}
	{d\over dt} \, x_{i}
	&=&
	    \left( f^a_{i j} \, x_a \right) H^{(j)}
	+
	    \left( f^a_{i j} \, g_{a k} + f^a_{i k} \, g_{j a} \right) H^{(j k)}
	+
	    x_a \left( f^a_{i j} \, g_{k l} +
				f^a_{i k} \, g_{j l} +
				f^a_{i l} \, g_{j k}
			\right)
	    H^{(j k l)}
	+ {\cal O}(\chi^2) \,,
    \\
    \noalign {\hbox{and}}
	{d\over dt} \, g_{i j}
	&=&
	    \left( f^a_{i k}\, g_{a j} + f^a_{j k}\, g_{i a} \right) H^{(k)}
	+
	    x_a \left( f^a_{i k}\, g_{j l} +
				f^a_{i l}\, g_{k j} +
				f^a_{j k}\, g_{i l} +
				f^a_{j l}\, g_{k i}
			\right)
	    H^{(k l)}
	+ {\cal O}(\chi^2) \,.
    \end {eqnarray}%
    \end {mathletters}
    }

Given these equations of motion for the connected expectations of generators,
one can use (\ref{power-series-expansion-symbol})
to describe the dynamics of any classical operator in terms of its
symbol.
If $\hat V = \hat V(\{ x_i \} )$ is a (time-independent)
function of the generators, then
its time-dependent expectation value,
at next-to-next-to-leading order,
is given by
\begin{eqnarray}
    \langle \hat V(t) \rangle &=&
    \left.
    \left\{
	\bar V
	+ g_{i j}(t) \, V^{(i j)}
	+ g_{i j k}(t) \, V^{(i j k)}
	+ g_{i j k l}(t) \, V^{(i j k l)}
    \right\}
    \right|_{x_m = x_m(t)}
    + {\cal O}(\chi^3) \,,
\end{eqnarray}
where $x_i(t)$, $g_{i j}(t) = s_{i j}(t)$, and
$g_{i j k}(t) = s_{i j k}(t)$ are to be obtained
by integrating Eq's.~(\ref{x-evolution})--(\ref{xxx-evolution})
forward in time, using
$
    g_{i j k l}(t) =
    g_{i j}(t) \, g_{k l}(t) +
    g_{i k}(t) \, g_{j l}(t) +
    g_{i l}(t) \, g_{j k}(t) +
    {\cal O}(\chi^3)
$.

\subsection{Error Accumulation} \label{errors}

To any given order in $\chi$,
we have a system of non-linear, first-order, ordinary
differential equations.
Initial conditions are imposed by specifying
$x_i(t{=}0) = \langle u | \hat x_i | u \rangle$
and $s^{(k)}_{j \cdots l}(t{=}0) =
\langle u | \hat x_j \cdots \hat x_l | u \rangle^{\rm conn}$,
with 
$|u\rangle$ some chosen coherent state.
Since $s^{(k)}(t{=}0)$ is ${\cal O}(\chi^{k-1})$,
and the equations for $\dot {s}^{(k)}(t)$ involve only
terms of order $\chi^{k-1}$ and higher, we still formally have
$s^{(k)}(t) = {\cal O}(\chi^{k-1})$ for $t > 0$.
However, as the truncated equations of motion
are integrated forward in time, errors accumulate;
it is important to understand the rate of growth of this truncation error.

We are dealing with a system of equations which we can write as
$ \dot y_i = F_i(y) + G_i(y)$
where $\{ y_i(t) \}$ are the variables in our problem
(that is, the $x_i$'s and $s^{(k)}_{\cdots}$'s),
$F(y)$ represents the terms we keep,
and $G(y)$ stands for everything thrown away by the truncation.
Let $y_0(t)$ be the solution to the above
equation with $G\equiv0$, and solve perturbatively,
$y(t) = y_0(t) + \epsilon(t)$
with $\epsilon$ small.
Linearizing about $y(t) = y_0(t)$, we have
\begin {equation}
    \dot \epsilon = f(t) \, \epsilon + g(t) \,,
\label {eq:trunc}
\end {equation}
where $f_i^j(t) = \partial F_i(y_0(t)) / \partial y_j$,
$g_i(t) = G_i(y_0(t))$,
and we have dropped ${\cal O}(\epsilon^2)$ terms.
This linearized system of equations is easy to solve
(at least formally).
For $t > 0$,
\begin {equation}
    \epsilon(t) =
    \left[ {\cal T} \, \e^{\int_0^t f(t') \> dt'} \right]
    \epsilon(0) +
    \int_0^t \left[ {\cal T} \, \e^{\int_{t'}^{t} f(t'') \> dt''} \right]
    g(t') \> dt' \,.
\end {equation}
Here, ${\cal T}$ denotes time ordering
(with smaller times on the right).
If $f(t)$ and $g(t)$ are globally bounded during the time evolution,
$|| f(t) || \le \tilde f$, $|| g(t) || \le \tilde g$,
where $||\cdots||$ is some appropriate norm,
then a crude estimate of the deviation of the true solution
from the approximation is
\begin{equation}
    \label{crude bound}
    || \epsilon(t) ||
    \le
    \e^{\tilde f t} \, || \epsilon(0) ||
    + \tilde g \, 
    {(\e^{\tilde f t} {-} 1 ) / \tilde f} \,.
\end{equation}
Of course for $t$ small, errors grow linearly and
$|| \epsilon(t) || \le 
|| \epsilon(0) || \, (1 {+} \tilde f t) + \tilde g t + {\cal O} (t^2)$; 
with a truncation good to order $\chi^k$ at $t=0$,
both $||\epsilon(0)||$ and $\tilde g$ will be ${\cal O}(\chi^{k+1})$.

In a general treatment, it is hard to do better than the
crude bound (\ref{crude bound}). 
In dynamical systems with only a few degrees of freedom,
there typically are ``regular'' portions of phase space where
perturbations grow only linearly with time \cite {Arnold}.
However, it is not at all clear that this is applicable to the
truncated quantum dynamics represented by Eq.~(\ref {eq:trunc}).

In simple examples discussed in the following section,
we will find that for times of order $\chi^{-1/2}$,
the shape of the wavefunction of the evolving state becomes
so distorted that the formal hierarchy of correlators,
$s^{(k)} \sim {\cal O}(\chi^{k-1})$,
upon which the truncation scheme is based,
completely breaks down.
In terms of the underlying quantum dynamics,
if one considers the projection of the initial coherent state
wavepacket
onto the exact eigenstates of the Hamiltonian, what is happening
for sufficiently large time
is that the contributions of different eigenstates have decohered
to such an extent that the wavepacket has spread beyond recognition.
Except for special non-generic cases 
(such as the harmonic oscillator, where there is no dispersion)
one should always expect such decoherence to eventually set in.

\section{Examples}

We will discuss two examples of theories to which the preceding
general results may be applied:
the usual semi-classical limit of point particle quantum mechanics,
and the large $N$ limit of $O(N)$ invariant vector models.
For brevity of presentation, we will display explicitly only
the first corrections to the leading classical approximation,
but we emphasize that it is completely straightforward to include
yet higher order corrections,
such as the ${\cal O}(\chi^2)$ terms displayed in
Eq's.~(\ref {x-evolution})--(\ref {xxx-evolution}).

\subsection {\boldmath$\hbar\to0$ Quantum Mechanics}
\label {ordinary QM}

Consider ordinary point particle quantum mechanics,
in one dimension for simplicity.
The coherence group $G$ is the
Heisenberg group, generated by
$\{e_i\} = \{\hat x / \hbar, \hat p / \hbar, {\bf 1} / \hbar \}$.
The formal parameter that controls how close the theory is to the
classical limit is, of course, $\chi = \hbar$.
The rescaled generators of the coherence group,
$\hat x_i = \hbar e_i$,
include the position $\hat x$ and momentum $\hat p$ operators
whose expectations will serve as classical phase space coordinates.
The Heisenberg group, acting on a fixed Gaussian base state,
generates conventional
coherent states $\{ | p, q \rangle \}$, with wave functions given
(up to an overall phase) by
\begin{equation}
\label {wf:QM}
\langle x | p, q \rangle =
(\pi \hbar)^{-1/4} \exp
\left\{ {1\over\hbar} \left[ i p x - \half (x{-}q)^2 \right] \right\}.
\end{equation}
We have arbitrarily chosen units such that our Gaussian base state has equal
variance in $\hat x$ and $\hat p$.
Consider a Hamiltonian of the typical form
$\hat H = {1\over2} \hat p^2 + V(\hat x)$,
where, for simplicity, we have set the particle mass to unity.
The equations of motion are, of course,
\begin {equation}
    {d \over dt} \, \hat x = \hat p \,, \qquad\qquad
    {d \over dt} \, \hat p = -V'(\hat x) \,.
\label {eq:EOMxp}
\end {equation}

We are interested in the time evolution of $x(t)$, $p(t)$, and
the connected correlators
$g_{xx}(t)$, $g_{xp}(t) = g^*_{px}(t)$, and
$g_{pp}(t)$, all to order $\hbar$. 
From equations (\ref{x-evolution}) and
(\ref{xx-evolution}) we find:
\begin{equation}
\begin {array}{rcll}
    \dot x & = & p
		    & {} + {\cal O} ( \hbar^2 ) \,,
\\
    \dot p & = & - V' - {1\over 2} V''' g_{x x}
		    &{} + {\cal O} ( \hbar^2 ) \,,
\\
    \dot{g}_{x x} &=& g_{x p} + g_{p x}
		    &{} + {\cal O} ( \hbar^2 ) \,,
\\
    (\dot{g}_{p x})^* = \dot{g}_{x p} &=& g_{p p} - V'' g_{x x}
		    &{} + {\cal O} ( \hbar^2 ) \,,
\\
    \dot{g}_{p p} &=& - V'' \left( g_{x p} + g_{p x} \right)
		    &{} + {\cal O} ( \hbar^2 ) \,,
\end {array}
\label{heisenberg-evolution}
\end{equation}
subject to the initial conditions
$x(0) = x_0$, $p(0) = p_0$,
$g_{x x} = g_{p p} = \half \hbar$,
$g_{x p} = - g_{p x} = \half i \hbar$.
Notice that to this order, 
$\det g^{(2)} = g_{xx} g_{pp} - g_{xp} g_{px} = {\cal O}(\hbar^3)$
is a constant of the motion, and Eq's.~(\ref{heisenberg-evolution}) are
equivalent to a Gaussian variational ansatz \cite{MDCBH}
(where one approximates the wave packet by a Gaussian
with a time-dependent centroid and width). However, if we went to
next-to-next-to-leading order in $\hbar$
it would become clear that our setup is different.
For positive times, higher moments will not be given by simple
algebraic expressions in terms of the centroid and variance, and
the details of evolution will depend on the shape of the potential.%
\footnote
    {
    In our Gaussian initial state, all connected correlators
    higher than second order vanish at time zero,
    $s^{(k > 2)}(0) \equiv 0$.
    But these moments cannot remain zero unless the potential is harmonic.
    For example, using Eq. (\ref{xxx-evolution}) we find that
    $\dot s_{xpx} =  s_{ppx} + s_{xpp} - V'' s_{xxx}
    - V''' \left( s_{xx} \right)^2 + {\cal O}(\hbar^3)$,
    showing explicitly that any nonzero $V'''$ will drive
    the skewness moments $s_{ijk}(t)$ away from zero.
    }

As a trivial warm-up, consider the harmonic oscillator of unit mass
and natural frequency $\Omega$:
$\hat H = {1\over2} \hat p^2 +
{1\over2} \Omega^2 \hat x^2$.
The solutions to (\ref{heisenberg-evolution}) are
\begin {mathletters}
\begin {eqnarray}
    x(t) &=& x_0 \cos \Omega t + (p_0/\Omega) \sin \Omega t \,,
\\
    p(t) &=& p_0 \cos \Omega t - (x_0 \, \Omega) \sin \Omega t \,,
\\
    g_{x x}(t) &=& {\textstyle {\hbar \over 2}}
    \left[ \cos^2 \Omega t  + \Omega^{-2} \sin^2 \Omega t \right] ,
\\
    g^*_{p x}(t) =
    g_{x p}(t) &=& {\textstyle {\hbar \over 2}}
    \left[i + (\Omega^{-1} {-} \Omega) \cos \Omega t \sin \Omega t \right] ,
\\
    g_{p p}(t) &=& {\textstyle  {\hbar \over 2}}
    \left[ \cos^2 \Omega t  + \Omega^2 \sin^2 \Omega t \right] .
\end {eqnarray}%
\end {mathletters}
Because the potential is quadratic these are exact.
Equally simple is an inverted harmonic oscillator.
If one takes the Hamiltonian to be
$\hat H = {1\over2} \hat p^2 - {1\over2} \Omega^2 \hat x^2$,
then the solution of the moment equations (\ref{heisenberg-evolution}) becomes
\begin {mathletters}
\begin {eqnarray}
    x(t) & = & x_0 \cosh \Omega t + (p_0/\Omega) \sinh \Omega t \,,
\\
    p(t) & = & p_0 \cosh \Omega t + (x_0 \, \Omega) \sinh \Omega t \,,
\\
    g_{x x}(t) &=& {\textstyle {\hbar \over 2}}
    \left[ \cosh^2 \Omega t  + \Omega^{-2} \sinh^2 \Omega t \right] ,
\\
    g^*_{p x}(t) =
    g_{x p}(t) &=& {\textstyle {\hbar \over 2}}
    \left[i + (\Omega^{-1} {+} \Omega)
    \cosh \Omega t \sinh \Omega t \right] ,
\\
    g_{p p}(t) &=&  {\textstyle {\hbar \over 2}}
    \left[ \cosh^2 \Omega t  + \Omega^2 \sinh^2 \Omega t \right] .
\end {eqnarray}%
\end {mathletters}
In both of these examples, the time-evolution of
the variances are
independent of $x_0$ and $p_0$.
As one would expect, they 
oscillate (with twice the natural frequency) in the case of the
simple harmonic oscillator, and grow (exponentially) for the inverted
oscillator.

As a more complicated example, consider the problem of small
oscillations in a weakly anharmonic potential,%
\footnote
    {
    We choose the curvature of the potential at the minimum to equal unity,
    so that our chosen coherent states have the natural
    width for the unperturbed potential.
    This ensures that the resulting dynamics
    (such as oscillations of $g^{(2)}$) are
    not merely reflecting purely harmonic oscillations.
    \label{omega-equals-one}
    }
$V(x) = {1\over2} x^2 + \beta x^4$.
The moment equations
(\ref{heisenberg-evolution}) become
\begin{equation}
\label{anharmonic-eom}
\begin {array} {rcll}
    \ddot x & = & -x - 4 \beta x^3 - 12 \beta \, x \, g_{x x}
    &{} + {\cal O} ( \hbar^2 ) \,,
\\
    \dot{g}_{xx} &=& g_{x p} + g_{p x}
    &{} + {\cal O} ( \hbar^2 ) \,,
\\
    (\dot{g}_{px})^* = \dot{g}_{x p} &=&
    g_{pp} - g_{xx} - 12 \beta \, x^2 \, g_{x x} 
    &{} + {\cal O} ( \hbar^2 ) \,,
\\
    \dot{g}_{pp} &=&
    - (1 + 12 \beta x^2) \left( g_{xp} + g_{px} \right)
    &{} + {\cal O} ( \hbar^2 ) \,.
\end {array}
\end{equation}
We will solve these perturbatively; the two small parameters
are $\beta q^2$ and $\beta \hbar$.
We will work to first order in $\beta \hbar$ (since we have omitted
${\cal O}(\hbar^2)$ terms in the moment equations), and will display
explicit results through second order in $\beta q^2$.
In principle, one could work to any order in $\beta q^2$ desired.

In order to keep our error estimates simple, we will treat the time
as ${\cal O}(1)$ (in units where the natural frequency is unity).
This means we need not worry about the appearance of secular terms
--- terms which grow as powers of $t$ --- and
may solve Eq's.~(\ref{anharmonic-eom}) strictly  perturbatively
in the naive fashion.
A straightforward calculation,
with the initial conditions
$x(0) = q$, $p(0) = 0$,
$g_{x x} = g_{p p} = \half \hbar$,
and
$g_{x p} = g_{p x}^* = \half i \hbar$,
leads to the solution
\begin{eqnarray}
\label{x-anharmonic}
    x(t) &=& q \> \Biggl\{ \cos t 
	+ \left(\textstyle {1\over 8} \beta q^2 \right)
	    \left[ \cos 3 t - \cos t - 12 t \sin t\vphantom{^0} \right]
\nonumber\\ && \kern .55in {}
	+ \left(\textstyle {1\over 8} \beta q^2 \right)^2
	    \left[
		\cos 5 t - 24 \cos 3 t + 23 \cos t 
		+ 96 \, t \sin t - 36 \, t \sin 3 t
		- 72 \, t^2 \cos t
	    \right]
\nonumber\\ &&  \kern .55in {}
    + (\beta \hbar)
	\left[
	    -3 \, t \sin t \vphantom{^0}
	\right]
\nonumber\\ &&  \kern .55in {}
    + (\beta \hbar)
      \left(\textstyle {1\over 8} \beta q^2 \right)
	\left[ 
	    -{\textstyle {15 \over 4}} (\cos 3t - \cos t)
	    - 18 \, t \sin 3 t + 93 \, t \sin t - 54 \, t^2 \cos t
	\right] 
\nonumber\\ && \kern .55in {}
	+ {\cal O}\!\left[ (\beta q^2)^3 \right]
	+ {\cal O}\!\left[ (\beta\hbar)(\beta q^2)^2 \right]
	+ {\cal O}\!\left[ (\beta \hbar)^2 \right]
    \Biggr\},
\end{eqnarray}
with 
\begin{eqnarray}
\label{g2-anharmonic}
    g_{xx}(t)
    &=& 
    \half \hbar
    \left\{
	1 - 3 \beta q^2 \left[1 - \cos 2 t + t \sin 2 t \vphantom{^0}\right]
	+ {\cal O}\!\left[(\beta q^2)^2 \right]
	+ {\cal O}(\beta \hbar)
    \right\} ,
\\
    g_{px}^*(t)
    = g_{xp}(t)
    &=& 
    \half \hbar
    \left\{
	i
	- {\textstyle {3 \over 2}} \, \beta q^2 
	    \left[ 3\sin 2 t + 2 t \cos 2 t \vphantom{^0}\right] ~
	+ {\cal O}\!\left[(\beta q^2)^2 \right]
	+ {\cal O}(\beta \hbar)
    \right\} ,
\\
    g_{pp}(t)
    &=& 
    \half \hbar
    \left\{
	1 + 3 \beta q^2 \left[1 - \cos 2 t + t \sin 2 t \vphantom{^0}\right]
	+ {\cal O}\!\left[(\beta q^2)^2 \right]
	+ {\cal O}(\beta \hbar)
    \right\} .
\end{eqnarray}
Examining the secular terms in Eq's.~(\ref {x-anharmonic}) and
(\ref {g2-anharmonic}), one sees that terms of order $\beta^k$
are accompanied by at most $k$ powers of $t$.
This is a general result.
It implies that our stated condition that the time be ${\cal O}(1)$ is
needlessly restrictive.
For small $\beta q^2$ and $\beta \hbar$,
the perturbative expansions (\ref {x-anharmonic}) and (\ref {g2-anharmonic})
are actually valid in the wider domain
$|\beta q^2 t| \ll 1$ and $|\beta \hbar t| \ll 1$,
provided a factor of $t$ is included with each factor of
$\beta q^2$ or $\beta \hbar$ in the error estimates.

It is instructive to compare this treatment with the result of a perturbative
quantum mechanical calculation.
Using the brute-force approach of first finding perturbed eigenstates
and energy levels, and then evaluating the time-dependent
expectation value $x(t)$ by projecting the initial coherent state
onto individual eigenstates and summing the resultant contributions,
a rather tedious calculation using both wavefunctions and
energies correct to ${\cal O}(\beta)$ leads to
\begin {eqnarray}
    x(t)
    &=&
    q \, \Biggl(
	e^{-(q^2/\hbar)\sin^2(3\beta\hbar t/2)} \,
	\cos\biggl[
	    (1 + 3 \beta\hbar) \, t + {q^2 \over 2\hbar} \sin (3\beta \hbar t)
	    \biggr]
\nonumber
\\ && {}
    +
	(\beta\hbar) \,
	e^{-(q^2/\hbar)\sin^2(3\beta\hbar t/2)}
	\Biggl\{
	    \cos\biggl[
	    (1{+}3 \beta\hbar) \, t + {q^2 \over 2\hbar} \sin (3\beta \hbar t)
	    \biggr]
	    \biggl(
		- {3 \over 2}
		- {q^2 \over \hbar}
		- {1 \over 32} \, {q^4 \over \hbar^2}
	    \biggr)
\nonumber
\\ && \kern 1.6in {}
    +
	    \cos\biggl[
	    (1{+}6 \beta\hbar) \, t + {q^2 \over 2\hbar} \sin (3\beta \hbar t)
	    \biggr]
	    \biggl(
		{3 \over 2}
		- {3\over 4} \, {q^2 \over \hbar}
		- {1 \over 4} \, {q^4 \over \hbar^2}
	    \biggr)
\nonumber
\\ && \kern 1.6in {}
    +
	    \cos\biggl[
	    (1{+}9 \beta\hbar) \, t + {q^2 \over 2\hbar} \sin (3\beta \hbar t)
	    \biggr]
	    \biggl(
		{3\over 2} \, {q^2 \over \hbar}
	    \biggr)
\nonumber
\\ && \kern 1.6in {}
    +
	    \cos\biggl[
	    (1{+}12 \beta\hbar) \, t + {q^2 \over 2\hbar} \sin (3\beta \hbar t)
	    \biggr]
	    \biggl(
		{1\over 8} \, {q^2 \over \hbar}
		+ {1 \over 4} \, {q^4 \over \hbar^2}
	    \biggr)
\nonumber
\\ && \kern 1.6in {}
    +
	    \cos\biggl[
	    (1{+}15 \beta\hbar) \, t + {q^2 \over 2\hbar} \sin (3\beta \hbar t)
	    \biggr]
	    \biggl(
		{1\over 32} \, {q^4 \over \hbar^2}
	    \biggr)
	\Biggr\}
\nonumber
\\ && {}
    +
	(\beta\hbar) \,
	e^{-(q^2/\hbar)\sin^2(9\beta\hbar t/2)} \,
	\cos\biggl[
	  3\,(1{+}6 \beta\hbar) \, t + {q^2 \over 2\hbar} \sin (9\beta \hbar t)
	    \biggr]
	    \biggl(
		{1\over 8} \, {q^2 \over \hbar}
	    \biggr) 
\Biggr) \,.
\label {eq:QM}
\end {eqnarray}
This result has ${\cal O}(\beta^2)$ errors due to the neglect of
second (and higher) order corrections in both the eigenstates
and energy eigenvalues.

If one restricts $t$ to be small compared to both $1/|\beta\hbar|$
and $1/|\beta q^2|$,
then one may expand the result (\ref {eq:QM}) in powers of $\beta$.
Moreover, in this domain one may easily add in the leading secular
${\cal O}[(\beta \hbar)^2 t]$ terms omitted from (\ref {eq:QM}),
which come from including the ${\cal O}(\beta^2)$ perturbation to
energy levels while using unperturbed wavefunctions.%
\footnote
    {%
    This addition is
    $
	q \, (t \, \sin t)
	\left[
	    {51\over16} \, (\beta q^2)^2
	    + {153\over8} \, (\beta q^2)(\beta \hbar)
	    + 18 (\beta \hbar)^2
	\right]
    $.
    If one does not assume that $\beta\hbar t$ is small compared to 1,
    then including the ${\cal O}(\beta^2)$ energy shift in matrix elements of
    time-evolution operators unfortunately leads to an analytically
    intractable infinite sum for $x(t)$.
    }
One finds
\begin{eqnarray}
    x(t) &=& q \Biggl\{ \cos t 
	+ \left(\textstyle {1\over 8} \beta q^2 \right)
	    \left[ \cos 3 t {-} \cos t {-} 12 t \sin t\vphantom{^0} \right]
	+ \left(\textstyle {1\over 8} \beta q^2 \right)^2
	    \left[
		96 \, t \sin t {-} 36 \, t \sin 3 t {-} 72 \, t^2 \cos t
	    \right]
\nonumber\\ &&  \kern .55in {}
    + (\beta \hbar)
	\left[
	    -3 \, t \sin t \vphantom{^0}
	\right]
    + (\beta \hbar)
      \left(\textstyle {1\over 8} \beta q^2 \right)
	\left[ 
	    - 18 \, t \sin 3 t + 93 \, t \sin t - 54 \, t^2 \cos t
	\right] 
\nonumber\\ && \kern .55in {}
	+ {\cal O}\!\left[ (\beta q^2{+}\beta\hbar)^2 \right]
	+ {\cal O}\!\left[ (\beta q^2{+}\beta\hbar)^3 \, t^3 \right]
    \Biggr\}.
\label {eq:check}
\end{eqnarray}
This result is perfectly consistent with the
previous moment-hierarchy result (\ref {x-anharmonic}),
as it must be, except for the non-secular ${\cal O}(\beta^2)$
terms which are hiding in the first
${\cal O}\!\left[ (\beta q^2{+}\beta\hbar)^2 \right]$ error term
of (\ref {eq:check}).
If one includes second order perturbations to
the eigenstates then these terms also coincide.

In the semi-classical regime, where $\beta \hbar \ll \beta q^2$,
it is interesting to examine expression (\ref{eq:QM}) when
$\beta\hbar t \ll 1$, making no assumption about the size of $\beta q^2 t$.
In this regime, the first, leading term of Eq.~(\ref {eq:QM})
becomes
\begin {equation}
    x(t)
    =
    q \, 
	e^{-{9 \over 4}(\beta q^2)(\beta\hbar) \, t^2} \,
	\cos\left[
	    (1 + 3 \beta\hbar + {\textstyle {3\over2}} \beta q^2) \, t \,
	    \right]
	+ {\cal O}(q\beta\hbar) \, .
\end {equation}
In other words, $x(t)$ shows damped harmonic behavior, with a
shifted $q$-dependent frequency, and with an amplitude which
decays significantly on the time scale
\begin {equation}
    t_d \sim \left[ (\beta q^2)(\beta \hbar) \right]^{-1/2} \,.
\label {eq:decoher}
\end {equation}
This implies that
on this time scale, the initially well localized wavepacket
has dispersed so much that its probability distribution is spread out
over most of the classically allowed region.%
\footnote
    {%
    Of course, the fact that the amplitude of oscillations in
    the mean position $x(t)$ decays on the decoherence time scale $t_d$
    cannot mean that the wavepacket has come to rest at the bottom
    of the potential while remaining a well-localized wavepacket,
    as this would violate energy conservation.
    In the semi-classical regime under discussion,
    the position of the initial wavepacket is significantly displaced
    from the minimum of the potential, $q^2 \gg \hbar$, implying that
    the total energy is large compared to the zero-point energy.
    Therefore, a negligible mean position at large times necessarily
    indicates that the wavepacket has spread so much that its probability
    density, at any late time, is delocalized over the entire classically
    allowed region, and no longer ``sloshes'' back-and-forth
    to any significant extent.
    Within the classically allowed region, energy conservation implies
    that the wavefunction must have substantial variations on scales
    far smaller than the (square root of the) variance in position ---
    which will be comparable to the width of the classically allowed region.
    }
Hence, $t_d$ should be regarded as a ``delocalization'' or ``decoherence'' time.
The higher order terms in Eq.~(\ref {eq:QM}) all exhibit essentially the same
behavior in this regime; each term oscillates with a (slightly different)
frequency and has an amplitude which decays on the decoherence time scale~$t_d$.

Although it will have no bearing on our discussion, it is interesting
to note that on yet longer time scales, when $t$ is near $2\pi/(3\beta\hbar)$
or integer multiplies thereof,
the exponential factors in Eq.~(\ref {eq:QM}) return to near unity,
implying that the time-dependent state has ``reassembled'' itself
into a recognizable wavepacket oscillating in the potential.%
\footnote
    {
    Whether this ``reassembly'' persists in the exact solution,
    or is an artifact of our first order perturbative result,
    is not entirely clear to us.
    }
Presumably, this is a reflection of the fact that this is an
integrable single degree of freedom system.

The existence of the decoherence time scale (\ref {eq:decoher})
has important consequences for the utility of any truncated
moment expansion, such as Eq's.~(\ref {x-evolution}--\ref {xxx-evolution}).
If the wavepacket has spread to such an extent that it is significantly
sampling all of its classically allowed region,
while necessarily retaining structure on smaller scales,
then the formal hierarchy of connected correlators,
$s^{(k)} \sim \hbar^{k-1}$, will have broken down.
Higher order moments will not be small compared to lower order ones.
Consequently, the moment expansion presented in
the previous section can only be useful for times which are small
compared to the decoherence time $t_d$.

The $1/\sqrt\hbar$ dependence of the decoherence time (\ref {eq:decoher})
may also be seen in another very simple example.
Consider the free evolution of a coherent state in the absence of any
potential.
As is well known, the width of the wavepacket grows without bound.
The evolution equations for $\hat x$ and $\hat p$ are, of course,
trivial,
$\hat p(t) = \hat p(0)$, and
$\hat x(t) = \hat x(0) + \hat p(0) \, t $.
Hence,
$
    \hat x^2(t) = \hat x^2(0)
    + [\hat x(0) \hat p(0) + \hat p(0) \hat x(0)] \, t
    + \hat p(0)^2 \, t^2
$, 
and so for our initial Gaussian coherent state
(with equal variance in $x$ and $p$),
\begin {eqnarray}
    g_{xx}(t)
    &=& g_{xx}(0) + [g_{xp}(0)+g_{px}(0)] \, t + g_{pp}(0) \, t^2 
\nonumber\\
    &=& \half \hbar \, (1+t^2) \,.
\end {eqnarray}
Here also,
we see that for times of order $\hbar^{-1/2}$
the hierarchy of correlators $s^{(k)}(t) \sim {\cal O}(\hbar^k)$
no longer holds.

We believe this to be a general result.
Whenever a semiclassical system exhibits dispersion,
the decoherence time is expected to scale as $\hbar^{-1/2}$,
and truncations of the moment hierarchy equations
(\ref {x-evolution}--\ref {xxx-evolution}) will only be accurate
for times small compared to the decoherence time.

\subsection{Vector Models}

Consider an $O(N)$ invariant theory whose fundamental degrees of
freedom form $O(N)$ vectors.
For simplicity, we will assume that the degrees of freedom are all
bosonic,%
\footnote
    {
    Extending this discussion to $U(N)$ invariant fermionic models
    is completely straightforward.
    }
and divided into a set of canonical coordinates
$\{ \hat x^i_\alpha \}$ and corresponding canonical momenta
$\{ \hat p^j_\beta \}$.
Here $i, j = 1, \cdots , N$ are $O(N)$ vector indices, while
$\alpha, \beta = 1, \cdots , m$ distinguish different $O(N)$ vectors.
These basic operators are assumed to satisfy canonical commutation
relations, normalized such that
$
    [ \hat x^i_\alpha, \hat p^j_\beta ]
    =
    (i / N) \, \delta^{ij} \, \delta_{\alpha\beta}
$.
In other words, we have chosen to scale both coordinates and moments
by $1/\sqrt N$ compared to their textbook form.
The small parameter controlling the approach to the classical limit
is $\chi \equiv 1/N$;
$\hbar$ has been set to unity.
The Hamiltonian is assumed to be $O(N)$ invariant, and we will completely
restrict attention to the $O(N)$ invariant sector of the theory.
Consequently, the relevant Hilbert space
${\cal H}_N$ is the space of all $O(N)$ invariant states, and all
physical operators can be constructed from the basic bilinears
\begin{mathletters}
\begin{eqnarray}
    \hat A_{\alpha\beta}
    &\equiv&
    \sum_{i=1}^N \> \hat x^i_\alpha \, \hat x^i_\beta \,
    ,
\\
    \hat B_{\alpha\beta}
    &\equiv&
    \sum_{i=1}^N \> \half \,\{
    \hat x^i_\alpha \, \hat p^i_\beta + \hat p^i_\beta \, \hat x^i_\alpha \} ,
\\
    \hat C_{\alpha\beta}
    &\equiv&
    \sum_{i=1}^N \> \hat p^i_\alpha \, \hat p^i_\beta \,.
\end{eqnarray}%
\label{a-b-and-c}%
\end{mathletters}
It will be convenient to regard $\hat x^i_\alpha$ and $\hat p^i_\alpha$
as the components of $m \times N$ matrices, so that
the basic bilinears (\ref{a-b-and-c}) may be assembled into
$m \times m$ matrices,
\begin {equation}
    \hat A = \hat x \hat x^T \,,
    \qquad
    \hat B = \hat x \hat p^T - {\textstyle{i \over 2}} \, \hat \id \,,
    \quad\hbox{and}\quad
    \hat C = \hat p \hat p^T \,.
\label {eq:ABC}
\end {equation}
Viewed as matrices, $\hat A$ and $\hat C$ are symmetric,
while $\hat B$ is non-symmetric.
The individual components of $\hat A$, $\hat B$, and $\hat C$ are all
Hermitian operators acting on ${\cal H}_N$.

We will take the Hamiltonian to have the general form
\begin {equation}
    N \hat H_N = N \left[ \half \, \tr \hat C + V(\hat A) \right] .
\label {eq:H_N}
\end {equation}
The overall factor of $N$ (given our scaling of coordinates
and momenta by $1/\sqrt N$) is exactly what is needed to ensure
that the $N\to\infty$ limit is a classical limit in the framework
of section II.
The potential energy function $V(A)$ may be any chosen scalar-valued function
of a symmetric matrix $A$.
The kinetic energy takes the simple form
$
    \half \, \tr \hat C = \half \sum_{i,\alpha} (\hat p^i_\alpha)^2
$
if all degrees of freedom are scaled to have unit mass.
Two specific examples in this class of models are:
\begin {list}{}{\leftmargin 0pt\labelwidth 10pt \itemindent 15pt}
\item [$i$)]
    A single particle moving in a central potential in $N$-dimensions.
    This is the simplest possible example;
    the theory has only a single $O(N)$ coordinate vector
    [{\em i.e.}, $m = 1$].
    The Hamiltonian is
    \begin {equation}
	N \hat H_N
	=
	    N \left[ \half \vec p \cdot \vec p + V(\vec x \cdot \vec x) \right]
	=
	    N \left[ \half \hat C + V(\hat A) \right],
    \label {eq:single-vector}
    \end {equation}
    where $V(r^2)$ is now a function of just a one variable.%
    \footnote
	{
	In terms of coordinates and momenta which have
	not been rescaled by $N^{-1/2}$, 
	$
	    N \hat H_N
	    = \half \, \vec p^{\,2} +
	    N \, V({1\over N} \vec x^{\,2} ) \,
	$.
	}
\item [$ii$)]
    An $O(N)$-invariant $\phi^4$ field theory.
    The theory, defined on a spatial lattice,
    has field operators $\hat\phi^i_s$
    and conjugate momenta $\hat\pi^i_s$, where $s$ labels the sites
    of some $d$-dimensional lattice.
    The canonical commutation relations
    (after scaling $\hat\phi$ and $\hat\pi$ by $1/\sqrt N$)
    are
    $
	\left[ \hat\phi^i_s, \hat\pi^j_{s'}\right]
	= (i/ N) \, \delta^{ij} \, \delta_{ss'}
    $,
    and the quantum Hamiltonian is
    \begin {mathletters}
    \begin {eqnarray}
	N \hat H_N
	&=&
	N \sum_s \left[
		\half \, \hat \pi_s \cdot \hat \pi_s
		+ \half \, \vec\nabla \hat\phi_s \cdot \vec\nabla \hat\phi_s
		+ \half \mu^2 \, \hat\phi_s \cdot \hat\phi_s
		+ {\textstyle {\lambda \over 4}}
		    \left(\hat\phi_s \cdot \hat\phi_s \right)^2
		\right]
    \\ &=&
	N \sum_s \left\{
		\half \, \hat C_{ss}
		+ \half \left.\left[ (-\nabla_s^2 + \mu^2 ) \hat A_{ss'}
		    \right] \right|_{s'=s}
		+ {\textstyle {\lambda \over 4}} \left(\hat A_{ss} \right)^2
	\right\} .
    \end {eqnarray}%
    \label {eq:Hlattice}
    \end {mathletters}
    [Here $\nabla$ is a lattice forward difference operator,
    dot products denote the implicit sum over $O(N)$ indices,
    and factors of lattice spacing are suppressed for simplicity.]
    The number $m$ of $O(N)$ vectors [or the dimension of the matrices
    $\hat A$, $\hat B$, and $\hat C$] equals the total number of lattice sites.
    Ignoring the obvious notational changes ($x \to \phi$, $p \to \pi$),
    this theory has precisely the stated form of
    Eq's.~(\ref {eq:ABC})--(\ref {eq:H_N}).
    The lattice theory may, of course, be viewed as a natural discretization
    of the formal continuum theory where the field operators
    $\hat\phi^i(x)$ and $\hat\pi^i(x)$
    depend on continuous spatial coordinates and
    \begin {equation}
	N \hat H_N
	=
	N \int (d^dx) \left[
		\half \, \hat \pi(x) \cdot \hat \pi(x)
		+ \half \, \vec\nabla \hat\phi(x) \cdot \vec\nabla \hat\phi(x)
		+ \half \mu^2 \, \hat\phi(x) \cdot \hat\phi(x)
		+ {\textstyle {\lambda \over 4}}
		    \left(\hat\phi(x) \cdot \hat\phi(x) \right)^2
	    \right].
    \end {equation}
\end {list}

Returning to the general discussion,
a straightforward calculation shows that the commutators of the basic
bilinears are
\begin {mathletters}
\label{commutator-A-and-B}
\begin{eqnarray}
    {\textstyle {N\over i}}\, [ \hat A_{\alpha\beta} , \hat A_{\gamma\delta} ]
    &=&
    {\textstyle {N\over i}}\, [ \hat C_{\alpha\beta} , \hat C_{\gamma\delta} ]
    =
    0 \,,
\\
    {\textstyle {N\over i}}\, [ \hat A_{\alpha\beta} , \hat B_{\gamma\delta} ]
    &=&
    \hat A_{\alpha\gamma} \, \delta_{\beta\delta} +
    \hat A_{\beta\gamma} \, \delta_{\alpha\delta} \,,
\\
    {\textstyle {N\over i}}\, [ \hat B_{\alpha\beta} , \hat B_{\gamma\delta} ]
    &=&
    \hat B_{\gamma\beta} \, \delta_{\alpha\delta} -
    \hat B_{\alpha\delta} \, \delta_{\gamma\beta} \,,
\\
    {\textstyle {N\over i}}\, [ \hat A_{\alpha\beta} , \hat C_{\gamma\delta} ]
    &=&
    \hat B_{\alpha\gamma} \, \delta_{\beta\delta} +
    \hat B_{\beta\gamma} \, \delta_{\alpha\delta} +
    \hat B_{\alpha\delta} \, \delta_{\beta\gamma} +
    \hat B_{\beta\delta} \, \delta_{\alpha\gamma} \,,
\\
    {\textstyle {N\over i}}\, [ \hat B_{\alpha\beta} , \hat C_{\gamma\delta} ]
    &=&
    \hat C_{\beta\gamma} \, \delta_{\alpha\delta} +
    \hat C_{\beta\delta} \, \delta_{\alpha\gamma} \,.
\end{eqnarray}%
\end{mathletters}
In other words, the commutators of $\hat A$, $\hat B$, and $\hat C$
(as well as just $\hat A$ and $\hat B$) close and these operators
generate a Lie algebra.%
\footnote
    {
    The Lie algebra structure constants
    are
    $
	f_{A_{\alpha\beta} B_{\gamma\delta}}^{A_{\mu\nu}} =
	\half\left(
	\delta_{\mu\alpha} \delta_{\nu\gamma} \delta_{\beta\delta} +
	\delta_{\mu\beta} \delta_{\nu\gamma} \delta_{\alpha\delta} +
	\delta_{\nu\alpha} \delta_{\mu\gamma} \delta_{\beta\delta} +
	\delta_{\nu\beta} \delta_{\mu\gamma} \delta_{\alpha\delta}
	\right)
    $,
    $
	f_{B_{\alpha\beta} B_{\gamma\delta}}^{B_{\mu\nu}} =
	\delta_{\mu\gamma} \delta_{\nu\beta} \delta_{\alpha\delta} -
	\delta_{\mu\alpha} \delta_{\nu\delta} \delta_{\gamma\beta}
    $,
    $
	f_{A_{\alpha\beta} C_{\gamma\delta}}^{B_{\mu\nu}} =
	\delta_{\mu\alpha} \delta_{\nu\gamma} \delta_{\beta\delta} +
	\delta_{\mu\beta} \delta_{\nu\gamma} \delta_{\alpha\delta} +
	\delta_{\mu\alpha} \delta_{\nu\delta} \delta_{\beta\gamma} +
	\delta_{\mu\beta} \delta_{\nu\delta} \delta_{\alpha\gamma}
    $,
    and
    $
	f_{B_{\alpha\beta} C_{\gamma\delta}}^{C_{\mu\nu}} =
	\half\left(
	\delta_{\mu\beta} \delta_{\nu\gamma} \delta_{\alpha\delta} +
	\delta_{\mu\beta} \delta_{\nu\delta} \delta_{\alpha\gamma} +
	\delta_{\nu\beta} \delta_{\mu\gamma} \delta_{\alpha\delta} +
	\delta_{\nu\beta} \delta_{\mu\delta} \delta_{\alpha\gamma}
	\right)
    $,
    plus those trivially related by antisymmetry; all others vanish.
    The resulting Lie algebra of operators
    $
	\{ \hat \Lambda (a,b,c)
	\equiv
	i N \sum_{\alpha\beta}
	    (
		a_{\alpha\beta} \, \hat A_{\beta\alpha} +
		b_{\alpha\beta} \, \hat B_{\beta\alpha} +
		c_{\alpha\beta} \, \hat C_{\beta\alpha}
	    )
	\}
    $
    is isomorphic to the $sp(2m)$ algebra represented by
    the $2m$-dimensional matrices
    $
	\{ \lambda(a,b,c) \equiv
	\left[ {-b \atop a} \; {c \atop b^T} \right] \}
    $,
    where $b = ||b_{\alpha\beta}||$, {\em etc}.,
    and $a$ and $c$ are symmetric.
    }
The appropriate coherence group which will create suitable $O(N)$ invariant
coherent states may be taken to be the group generated by (anti-Hermitian
linear combinations of) the operators $\{ \hat A_{\alpha\beta} \}$ and
$\{ \hat B_{\alpha\beta} \}$.
Enlarging the coherence group by including the $\hat C_{\alpha\beta}$
operators among the generators is equally acceptable, but unnecessary.
The group generated by $\{ \hat A_{\alpha\beta} \}$ and
$\{ \hat B_{\alpha\beta} \}$
alone satisfies all the conditions for producing an over-complete
set of coherent states which behave classically as $N \to\infty$.
Including the $\hat C_{\alpha\beta}$ operators among the generators
enlarges the coherence group, but has no effect whatsoever on the
resulting manifold of coherent states.

Acting on an initial Gaussian base state, the coherence group generates
a set of coherent states $\{ |z\rangle \}$,
where $z$ is a complex symmetric $m\times m$ matrix,
with positive definite real part, which may be used to uniquely
label an individual coherent state.
The position space wavefunctions of these coherent states are given by
\begin {equation}
    \Psi_z(x) = 
    \det\left[{N \over 2\pi} \, (z + z^*)\right]^{N /4}
    \exp \left(
	-\half N \, \tr x^T z x
    \right) .
\label {eq:Psi}
\end {equation}
It will be convenient to decompose the matrix $z$
into its real and imaginary parts by writing
\begin {equation}
    z = \half a^{-1} - i \omega \,,
\label {eq:z}
\end {equation}
so that $a = (z {+} z^*)^{-1}$ and $\omega = {i\over2}(z {-} z^*)$.
Both $a$ and $\omega$ are real symmetric matrices,
and $a$ is positive definite.
Using the fact that
$
    \hat p^i_\beta |z\rangle
    =
    i \hat x^i_\alpha z_{\alpha\beta} |z\rangle
$,
a short exercise shows that the coherent state expectation values
of the basic bilinears are
\begin {equation}
    A(z) = a \,,\qquad
    B(z) = a \omega \,,\quad\hbox{and}\quad
    C(z) = \omega a \omega + {\textstyle {1\over 4}} \, a^{-1} \,.
\label {eq:ABCz}
\end {equation}
The variances of these operators in the coherent state $|z\rangle$ are%
\footnote
    {
    For aesthetic reasons, we set
    $
	g_{\alpha\beta, \gamma\delta}^{A B} \equiv
	g_{A_{\alpha\beta} B_{\gamma\delta}}
	= \langle A_{\alpha\beta} B_{\gamma\delta} \rangle
	- \langle A_{\alpha\beta} \rangle\langle B_{\gamma\delta} \rangle
    $,
    {\em etc.}
    }
\begin {mathletters}
\begin {eqnarray}
    g^{AA}_{\alpha\beta,\gamma\delta}
    &=&
    {\textstyle {1\over N}} \left[
	a_{\alpha\gamma} \, a_{\beta\delta} +
	a_{\alpha\delta} \, a_{\beta\gamma}
    \right],
\\
    g^{BB}_{\alpha\beta,\gamma\delta}
    &=&
    {\textstyle {1\over N}} \left[
	a_{\alpha\gamma} \, (\qtr a^{-1} {+} \omega a \omega)_{\beta\delta} +
	(\half + i \omega a)_{\beta\gamma} \,(\half - i a \omega)_{\alpha\delta}
    \right],
\\
    g^{CC}_{\alpha\beta,\gamma\delta}
    &=&
    {\textstyle {1\over N}} \left[ 
	(\qtr a^{-1} {+} \omega a \omega)_{\alpha\gamma} \,
	(\qtr a^{-1} {+} \omega a \omega)_{\beta\delta} +
	(\qtr a^{-1} {+} \omega a \omega)_{\alpha\delta} \,
	(\qtr a^{-1} {+} \omega a \omega)_{\beta\gamma}
    \right] \!,\!
\\
    \left( g^{BA}_{\gamma\delta,\alpha\beta} \right)^*
    =
    g^{AB}_{\alpha\beta,\gamma\delta}
    &=&
    {\textstyle {i \over N}} \left[
	a_{\alpha\gamma} \, (\half - i a\omega)_{\beta\delta} +
	a_{\beta\gamma} \, (\half - i a\omega)_{\alpha\delta}
    \right],
\\
    \left( g^{CA}_{\gamma\delta,\alpha\beta} \right)^*
    =
    g^{AC}_{\alpha\beta,\gamma\delta}
    &=&
    -{\textstyle {1\over N}} \left[
	(\half - i a\omega)_{\alpha\gamma} \,
	(\half - i a\omega)_{\beta\delta} +
	(\half - i a\omega)_{\alpha\delta} \,
	(\half - i a\omega)_{\beta\gamma}
    \right],
\\
    \left( g^{CB}_{\gamma\delta,\alpha\beta} \right)^*
    =
    g^{BC}_{\alpha\beta,\gamma\delta}
    &=&
    {\textstyle {i\over N}} \left[
	(\half - i a\omega)_{\alpha\gamma} \,
	(\qtr a^{-1} {+} \omega a \omega)_{\beta\delta} +
	(\half - i a\omega)_{\alpha\delta} \,
	(\qtr a^{-1} {+} \omega a \omega)_{\beta\gamma}
    \right].
\end {eqnarray}%
\end {mathletters}

Given our choice of Hamiltonian (\ref {eq:H_N}),
the operator equations of motion for the basic bilinears are
\begin {mathletters}
\begin {eqnarray}
    {d\over dt} \, \hat A_{\alpha\beta}
    &=&
    \hat B_{\alpha\beta} + \hat B_{\beta\alpha} \,,
\\
    {d\over dt} \, \hat B_{\alpha\beta}
    &=&
    \hat C_{\alpha\beta}
	- 2 \hat A_{\alpha\gamma} \, \hat V'_{\gamma\beta} \,,
\\
    {d\over dt} \, \hat C_{\alpha\beta}
    &=& 
    - 2 \hat B_{\gamma\alpha} \, \hat V'_{\gamma\beta}
    - 2 \hat V'_{\alpha\gamma} \, \hat B_{\gamma\beta} \,.
\end {eqnarray}%
\label{eq:ABCeom}
\end {mathletters}
Here, $V'$ is shorthand for the variation of
$V(A)$ with respect to the symmetric matrix $A$,
\begin {equation}
    V'_{\alpha\beta}
    \equiv
    {\delta V(A) \over \delta A_{(\alpha\beta)}}
    \equiv
    \half
    \left[
    {\delta V(A) \over \delta A_{\alpha\beta}} +
    {\delta V(A) \over \delta A_{\beta\alpha}}
    \right] ,
\end {equation}
and is defined so that
$\delta V(A) = {\rm tr} \> (V' \, \delta A)$.%
\footnote
    {%
    Note that with this definition, the matrix variation $V'$
    reduces to an ordinary variational derivative in the case of
    a single vector ($m{=}1$).
    }

Applying the general results (\ref {x-evolution}) and (\ref {xx-evolution})
[actually, only (\ref {eq:NLO}) is needed]
to the case at hand, one finds in a straightforward fashion
the following equations, valid to next-to-leading order in $1/N$,
for the time evolution of the expectation values and variances
of basic bilinears,
\begin {mathletters}
\begin {eqnarray}
    {d \over dt} \, A_{\alpha\beta}
    &=&
    B_{\alpha\beta} + B_{\beta\alpha}
    \,,
\\
    {d \over dt} \, B_{\alpha\beta}
    &=&
    C_{\alpha\beta}
    - 2 A_{\alpha\eta} \, V'_{\eta\beta}
    - 2 g^{AA}_{\alpha\eta,\mu\nu} \, V''_{\eta\beta,\mu\nu}
    - g^{AA}_{\mu\nu,\zeta\xi} \,
	A_{\alpha\eta} \, V'''_{\eta\beta,\mu\nu,\zeta\xi}
    + {\cal O}( N^{-2})
    \,,
\\
    {d \over dt} \, C_{\alpha\beta}
    &=&
    - 2 B_{\eta\alpha} \, V'_{\eta\beta}
    - 2 B_{\eta\beta} \, V'_{\eta\alpha}
    - \left( g^{BA}_{\eta\alpha,\mu\nu} + g^{AB}_{\mu\nu,\eta\alpha} \right)
	V''_{\eta\beta,\mu\nu}
    - \left( g^{BA}_{\eta\beta,\mu\nu} + g^{AB}_{\mu\nu,\eta\beta} \right)
	V''_{\eta\alpha,\mu\nu}
\nonumber\\ && {}
    - g^{AA}_{\mu\nu,\zeta\xi}
	\left(
	    B_{\eta\alpha} \, V'''_{\eta\beta,\mu\nu,\zeta\xi} +
	    B_{\eta\beta} \, V'''_{\eta\alpha,\mu\nu,\zeta\xi}
	\right)
    + {\cal O}(N^{-2})
    \,,
\end {eqnarray}%
\label {eq:ABCdot}%
\end {mathletters}
together with
\begin {mathletters}
\begin {eqnarray}
    {d \over dt} \, g^{AA}_{\alpha\beta,\gamma\delta}
    &=&
      g^{BA}_{\alpha\beta,\gamma\delta}
    + g^{BA}_{\beta\alpha,\gamma\delta}
    + g^{AB}_{\alpha\beta,\gamma\delta}
    + g^{AB}_{\alpha\beta,\delta\gamma}
    + {\cal O}( N^{-2}) \,,
\\
    {d \over dt} \, g^{BB}_{\alpha\beta,\gamma\delta}
    &=&
      g^{CB}_{\alpha\beta,\gamma\delta}
    + g^{BC}_{\alpha\beta,\gamma\delta}
    - 2 g^{AB}_{\alpha\eta,\gamma\delta} \, V'_{\eta\beta}
    - 2 g^{BA}_{\alpha\beta,\gamma\eta} \, V'_{\eta\delta}
\nonumber\\ && {}
    -\left( g^{AB}_{\mu\nu,\gamma\delta} +
	    g^{BA}_{\gamma\delta,\mu\nu}
     \right) A_{\alpha\eta} V''_{\eta\beta,\mu\nu}
\nonumber\\ && {}
    -\left( g^{AB}_{\mu\nu,\alpha\beta} +
	    g^{BA}_{\alpha\beta,\mu\nu} 
     \right) A_{\gamma\eta} V''_{\eta\delta,\mu\nu}
    + {\cal O}( N^{-2}) \,,
\\
    {d \over dt} \, g^{CC}_{\alpha\beta,\gamma\delta}
    &=&
    - 2 g^{BC}_{\eta\alpha,\gamma\delta} \, V'_{\eta\beta}
    - 2 g^{CB}_{\alpha\beta,\eta\gamma} \, V'_{\eta\delta}
    - 2 g^{BC}_{\eta\beta,\gamma\delta} \, V'_{\alpha\eta}
    - 2 g^{CB}_{\alpha\beta,\eta\delta} \, V'_{\gamma\eta}
\nonumber\\ && {}
    - \left( g^{AC}_{\mu\nu,\alpha\beta} + g^{CA}_{\alpha\beta,\mu\nu} \right)
	\left(
	    B_{\eta\delta} \, V''_{\gamma\eta,\mu\nu} +
	    B_{\eta\gamma} \, V''_{\eta\delta,\mu\nu}
	    \right)
\nonumber\\ && {}
    - \left( g^{AC}_{\mu\nu,\gamma\delta} + g^{CA}_{\gamma\delta,\mu\nu} \right)
	\left(
	    B_{\eta\beta} \, V''_{\alpha\eta,\mu\nu} +
	    B_{\eta\alpha} \, V''_{\eta\beta,\mu\nu}
	\right)
    + {\cal O}( N^{-2}) \,,
\\
    {d \over dt} \, g^{AB}_{\alpha\beta,\gamma\delta}
    &=&
      g^{BB}_{\alpha\beta,\gamma\delta}
    + g^{BB}_{\beta\alpha,\gamma\delta}
    + g^{AC}_{\alpha\beta,\gamma\delta}
    - 2 g^{AA}_{\alpha\beta,\gamma\eta} \, V'_{\eta\delta}
    - 2 g^{AA}_{\alpha\beta,\mu\nu} \, A_{\gamma\eta} \, V''_{\eta\delta,\mu\nu}
    + {\cal O}( N^{-2}) \,,
\\
    {d \over dt} \, g^{AC}_{\alpha\beta,\gamma\delta}
    &=&
      g^{BC}_{\alpha\beta,\gamma\delta}
    + g^{BC}_{\beta\alpha,\gamma\delta}
    - 2 g^{AB}_{\alpha\beta,\eta\delta} \, V'_{\gamma\eta}
    - 2 g^{AB}_{\alpha\beta,\eta\gamma} \, V'_{\eta\delta}
\nonumber\\ && {}
    - 2 g^{AA}_{\alpha\beta,\mu\nu}
	\left(
	    B_{\eta\delta} \, V''_{\gamma\eta,\mu\nu} +
	    B_{\eta\gamma} \, V''_{\eta\delta,\mu\nu}
	\right)
    + {\cal O}( N^{-2}) \,,
\\
    {d \over dt} \, g^{BC}_{\alpha\beta,\gamma\delta}
    &=&
      g^{CC}_{\alpha\beta,\gamma\delta}
    - 2 g^{AC}_{\alpha\eta,\gamma\delta} \, V'_{\eta\beta}
    - 2 g^{BB}_{\alpha\beta,\eta\delta} \, V'_{\gamma\eta}
    - 2 g^{BB}_{\alpha\beta,\eta\gamma} \, V'_{\eta\delta}
\nonumber\\ && {}
    - \left( g^{AC}_{\mu\nu,\gamma\delta} + g^{CA}_{\gamma\delta,\mu\nu} \right)
	    A_{\alpha\eta} \, V''_{\eta\beta,\mu\nu}
\nonumber\\ && {}
    - \left( g^{AB}_{\mu\nu,\alpha\beta} + g^{BA}_{\alpha\beta,\mu\nu} \right)
	\left(
	    B_{\eta\delta} \, V''_{\gamma\eta,\mu\nu} +
	    B_{\eta\gamma} \, V''_{\eta\delta,\mu\nu}
	\right)
    + {\cal O}( N^{-2}) \,.
\end {eqnarray}%
\label {eq:gdot}%
\end {mathletters}
Here
$
    V''_{\alpha\beta,\gamma\delta}
    \equiv
    { \delta^2 V(A) \over \delta A_{(\alpha\beta)} \delta A_{(\gamma\delta)} }
$,
{\em etc}.
Of course,
$g^{BA}_{\gamma\delta,\alpha\beta} = \left(g^{AB}_{\alpha\beta,\gamma\delta}\right)^*$
and so on,
since the basic bilinears $\hat A_{\alpha\beta}$, $\hat B_{\alpha\beta}$,
and $\hat C_{\alpha\beta}$ are all Hermitian.

As they stand,
the (truncated) moment equations (\ref {eq:ABCdot}) and (\ref {eq:gdot})
are highly redundant.
This is because the operators
$\hat A_{\alpha\beta}$, $\hat B_{\alpha\beta}$, and $\hat C_{\alpha\beta}$
are not independent
when acting on the $O(N)$ invariant Hilbert space ${\cal H}_N$.
For many purposes, it is preferable to reduce the evolution equations
to a smaller set of independent observables.
To see the redundancy, it is convenient first to note that the actions of
$\hat B$ and $\hat A$ on any coherent state $|z\rangle$ are related,
\begin {equation}
    \left(\hat B + \mbox{$i\over2$} \, \hat {\bf 1}\right) | z \rangle =
    \hat x \hat p^T | z \rangle = \hat x \hat x^T \, (iz) | z \rangle 
    =
    \hat A \, (iz) |z \rangle
    \,.
\end {equation}
Hence,
the coherent state expectation value of $\hat A^{-1} \hat B$ is
directly related to that of $\hat A^{-1}$,%
\footnote
    {%
    The following discussion assumes that coherent state matrix elements
    of $\hat A^{-1}$ exist, which requires $N > m+1$.
    }
\begin {equation}
    \langle z | \left( \hat A^{-1} \hat B \right)_{\alpha\beta} | z \rangle
    =
    i z_{\alpha\beta}
    - {\textstyle {i \over 2}} \,
    \langle z | \left( \hat A^{-1} \right)_{\alpha\beta} | z \rangle \,.
\label {eq:Bz}
\end {equation}
In a similar fashion, the 
coherent state expectation value of $\hat C$ may be expressed as
\begin{eqnarray}
    \langle z | \hat C | z \rangle
    &=&
    \langle z | \hat p \hat p^T | z \rangle
    =
    \langle z | (i z)^* \, (\hat x \hat x^T) \, (i z) | z \rangle
    =
    \langle z |
    (\hat p \hat x^T) \, (\hat x \hat x^T)^{-1} \, (\hat x \hat p^T)
    | z \rangle
\nonumber
\\ &=&
    \langle z |
    (\hat B + \mbox{$i\over2$} \hat {\bf 1})^\dagger \, \hat A^{-1} \,
    (\hat B + \mbox{$i\over2$} \hat {\bf 1})
    | z \rangle \,.
\label {eq:Cz}
\end{eqnarray}

As noted earlier in section II,
quantum operators are completely determined by their diagonal
expectation values in the over-complete coherent basis.
Consequently,
the coherent state relations
(\ref {eq:Bz}) and (\ref {eq:Cz}) suffice to infer
underlying operator identities.
The left-hand side of relation (\ref {eq:Bz}) is not manifestly symmetric
under interchange of $\alpha$ and $\beta$, but the right-hand side is
symmetric under this interchange.
Because Eq.~(\ref {eq:Bz}) holds for all coherent states $\{|z\rangle\}$,
if one defines
\begin {equation}
    \hat \V_{\alpha\beta}
    \equiv
    \left( \hat A^{-1} \hat B \right)_{\alpha\beta}
    + i \left({\textstyle {m{+}1 \over 2N}}\right)
	\left( \hat A^{-1} \right)_{\alpha\beta}
    \,,
\label {eq:Omega}
\end {equation}
then (\ref {eq:Bz}) implies that
$\V_{\alpha\beta} = \V_{\beta\alpha}$,
so that $\V = ||\V_{\alpha\beta}||$
is a symmetric matrix.
Moreover, using the the commutation relations (\ref{commutator-A-and-B}),
one may verify that $\V_{\alpha\beta}$ is Hermitian.
[Demanding Hermiticity is what determines the coefficient of the
second term in (\ref {eq:Omega}).]
Similarly, relation (\ref {eq:Cz}) implies the operator identity
\begin{equation}
    \hat C =
    \left( \hat B + {\textstyle {i\over2}} \hat {\bf 1} \right)^\dagger
    \hat A^{-1}
    \left( \hat B + {\textstyle {i\over2}} \hat {\bf 1} \right) ,
\label{c-in-terms-of-a-and-b}%
\end{equation}
showing that the operators $\{\hat C_{\alpha\beta}\}$ are not
independent of $\hat A$ and $\hat B$ [when acting on $O(N)$ invariant states].
Inverting the definition (\ref {eq:Omega}) to
express $\hat B$ in terms of $\hat \V$,
\begin {equation}
    \hat B = \hat A \, \hat \V
    - i \left({\textstyle {m{+}1 \over 2N}}\right) \hat \id \,,
\end {equation}
and using this, plus the Hermiticity of $\hat\V$,
allows one to rewrite expression (\ref {c-in-terms-of-a-and-b})
for $\hat C$ as
\begin {equation}
    \hat C
    =
    \hat \V \, \hat A \, \hat \V
    + {\textstyle {1\over 4}}
    \left( \textstyle 1 - {m{+}1 \over N}\right)^2
    \hat A^{-1} \,.
\end {equation}
Hence,
within the $O(N)$ invariant Hilbert space,
instead of working with the basic bilinears
$\hat A$, $\hat B$, and $\hat C$ [totaling $m(2m{+}1)$ distinct operators],
it is sufficient to use only $\hat A$ and $\hat \V$
[totaling $m(m{+}1)$ distinct operators].
These operators are, in fact, canonically conjugate
``coordinates'' and ``momenta''.
A short exercise shows that
\begin {mathletters}
\begin{eqnarray}
    \left[ \hat A_{\alpha\beta} , \hat A_{\gamma\delta} \right]
    &=&
    \left[ \hat \V_{\alpha\beta} , \hat \V_{\gamma\delta} \right]
    =
    0 \,,
\\
    i N
    \left[ \hat \V_{\alpha\beta} , \hat A_{\gamma\delta} \right]
    &=&
    \delta_{\alpha\gamma} \delta_{\beta\delta} +
    \delta_{\alpha\delta} \delta_{\beta\gamma}
    \,.
\end{eqnarray}%
\label{commutator-W-and-V}%
\end {mathletters}
If the complex symmetric matrix $z$ parameterizing coherent states
is separated into real and imaginary parts
by writing $z = \half a^{-1} - i \omega$
[as in Eq.~(\ref {eq:z})],
then the coherent state expectations of the canonical operators
$\hat A$ and $\hat \V$
are just $a$ and $\omega$, respectively,
\begin {equation}
    \langle z|\hat A|z\rangle = a, \qquad
    \langle z|\hat \V|z\rangle = \omega \,.
\end {equation}
[The first equality was previously noted in Eq.~(\ref {eq:ABCz}).]

Re-expressing the quantum equations of motion (\ref {eq:ABCeom})
in terms of the independent canonically conjugate operators gives
\begin {mathletters}
\begin{eqnarray}
    {d\over dt} \, \hat A &=& \hat A \, \hat \V + \hat \V \hat A \,,
\\
    {d\over dt} \, \hat \V &=& - \hat \V^2 - 2 \Veff'(\hat A) \,,
\end{eqnarray}
\label{large-N-quantum-evolution}%
\end{mathletters}
where the ``effective'' radial potential
\begin {equation}
    \Veff(A) \equiv V(A)
    + {\textstyle {1\over8}} \left(1 - {\textstyle {m+1\over N}} \right)^2
	\tr A^{-1}
\end {equation}
equals the original potential energy augmented by a 
``centrifugal potential''.

One may directly evaluate the evolution equations for
expectations and variances of the canonically conjugate operators
$\hat A$ and $\hat \V$,
or equivalently (and rather tediously) rewrite the previous equations 
(\ref {eq:ABCdot}) and (\ref {eq:gdot}) in terms of $\hat A$ and $\hat \V$.
Either way, one finds
\begin{mathletters}
\begin{eqnarray}
    {d \over dt} \, {A}_{\alpha\beta}
    &=&
    (A \V + \V A)_{\alpha\beta}
    + g_{\alpha\eta, \eta\beta}^{A \V}
    + g_{\alpha\eta, \eta\beta}^{\V A}
    + {\cal O}( N^{-2}) \,,
\\
    {d \over dt} \, {\V}_{\alpha\beta}
    &=&
    - (\V^2 + 2 \Veff' )_{\alpha\beta}
    - g_{\alpha\eta, \eta\beta}^{\V \V}
    - \left( \Veff'''\right)_{\alpha\beta,\mu\nu,\zeta\xi} \,
	g_{\mu\nu, \zeta\xi}^{A A}
    + {\cal O}( N^{-2}) \,,
\end {eqnarray}%
\label{large-N-evolution-expectationsA}
\end {mathletters}
together with
\begin {mathletters}
\begin {eqnarray}
    {d \over dt} \, {g}_{\alpha\beta, \gamma\delta}^{A A}
    &=&
    g_{\alpha\eta, \gamma\delta}^{A A}  \, \V_{\eta\beta} +
    g_{\eta\beta, \gamma\delta}^{\V A} \,  A_{\alpha\eta} +
    g_{\alpha\eta, \gamma\delta}^{\V A} \,  A_{\eta\beta} +
    g_{\eta\beta, \gamma\delta}^{A A}  \, \V_{\alpha\eta}
\nonumber
\\ && {}
    + g_{\alpha\beta, \gamma\eta}^{A A}  \, \V_{\eta\delta}
    + g_{\alpha\beta, \eta\delta}^{A \V}   \, A_{\gamma\eta}
    + g_{\alpha\beta, \gamma\eta}^{A \V} \, A_{\eta\delta}
    + g_{\alpha\beta, \eta\delta}^{A A}    \, \V_{\gamma\eta}
    + {\cal O}( N^{-2}) \,,
\\
    {d \over dt} \left( {g}_{\gamma\delta, \alpha\beta}^{\V A} \right)^*
    =
    {d \over dt} \, {g}_{\alpha\beta, \gamma\delta}^{A \V}
    &=&
    g_{\alpha\eta, \gamma\delta}^{A\V}  \, \V_{\eta\beta} +
    g_{\eta\beta, \gamma\delta}^{\V\V} \, A_{\alpha\eta} +
    g_{\alpha\eta, \gamma\delta}^{\V\V} \, A_{\eta\beta} +
    g_{\eta\beta, \gamma\delta}^{A\V}  \, \V_{\alpha\eta}
\nonumber
\\ && {}
    - g_{\alpha\beta, \eta\delta}^{A\V}   \, \V_{\gamma\eta}
    - g_{\alpha\beta, \gamma\eta}^{A\V} \, \V_{\eta\delta}
    - 2 \left( \Veff'' \right)_{\gamma\delta,\mu\nu} \,
	g_{\alpha\beta, \mu\nu}^{AA}
    + {\cal O}( N^{-2}) \,,
\\
    {d \over dt} \, {g}_{\alpha\beta, \gamma\delta}^{\V\V}
    &=&
    - g_{\alpha\beta, \eta\delta}^{\V\V}   \, \V_{\gamma\eta}
    - g_{\alpha\beta, \gamma\eta}^{\V\V} \, \V_{\eta\delta}
    - 2 \left( \Veff'' \right)_{\gamma\delta,\mu\nu} \,
	g_{\alpha\beta, \mu\nu}^{\V A}
\nonumber
\\ && {}
    - g_{\eta\beta, \gamma\delta}^{\V\V} \, \V_{\alpha\eta}
    - g_{\alpha\eta, \gamma\delta}^{\V\V} \, \V_{\eta\beta}
    - 2 \left( \Veff'' \right)_{\alpha\beta,\mu\nu} \,
	g_{\mu\nu, \gamma\delta}^{A\V}
    + {\cal O}( N^{-2}) \,.
\end{eqnarray}%
\label{large-N-evolution-expectationsB}%
\end{mathletters}%

Initial conditions corresponding to a given coherent
state $|z\rangle$ (with $z = \half a^{-1} {-}i\omega$)
are given by
$A(0) = a$ and $\V(0) = \omega$, together with the variances
\begin{equation}
\label{large-N-initial-conditions}
    \left(
	\matrix{
	    g_{\alpha\beta, \gamma\delta}^{A A} &
	    g_{\alpha\beta, \gamma\delta}^{A \V} \cr
	    g_{\alpha\beta, \gamma\delta}^{\V A} &
	    g_{\alpha\beta, \gamma\delta}^{\V \V}
	}
    \right)_{t=0}
    =
    {1\over N}
    \left(
	\matrix{
	    a_{\alpha\delta} a_{\beta\gamma} + a_{\alpha\gamma} a_{\beta\delta} &
	    {i\over2}
		[\delta_{\alpha\delta} \delta_{\beta\gamma} +
		\delta_{\alpha\gamma} \delta_{\beta\delta} ] \cr
	    - {i\over2}
		[\delta_{\alpha\delta} \delta_{\beta\gamma} +
		\delta_{\alpha\gamma} \delta_{\beta\delta} ] &
	    {1 \over 4}
		[a^{-1}_{\alpha\delta} a^{-1}_{\beta\gamma} +
		a^{-1}_{\alpha\gamma} a^{-1}_{\beta\delta} ]
	    }
    \right) 
    + {\cal O}(N^{-2})
    \,.
\end{equation}

The next-to-leading order evolution equations
(\ref {large-N-evolution-expectationsA})
and (\ref {large-N-evolution-expectationsB})
are directly applicable to any bosonic $O(N)$ invariant vector model,
such as the $\phi^4$ theory defined by (\ref {eq:Hlattice}),
whose Hamiltonian has the general form (\ref {eq:H_N}).
The dynamics is encoded in as efficient a form as possible;
one has dynamical equations
for the $m(m{+}1)/2$ pairs of independent phase space coordinates
(\ref {large-N-evolution-expectationsA}),
and their variances (\ref {large-N-evolution-expectationsB}).

In the special case (\ref {eq:single-vector})
of a single vector (corresponding to a point particle
moving in an $N$-dimensional spherically symmetric potential)
one may drop all the indices and the next-to-leading order
evolution equations become
\begin {mathletters}
\begin{eqnarray}
    {d \over dt} \, A
    &=&
    A \, \V + \V \, A
    + g_{A \V} + g_{\V A}
    + {\cal O} ( N^{-2} ) \,,
\\
    {d \over dt} \, \V
    &=&
    - \V^2 - 2\Veff' - g_{\V \V} - g_{A A} \, \Veff'''
    + {\cal O} ( N^{-2} ) \,,
\\
    {d \over dt} \, {g}_{A A}
    &=&
    4 g_{A A} \, \V + 2 (g_{\V A} + g_{A \V}) \, A
    + {\cal O} ( N^{-2} ) \,,
\\
    {d \over dt} \, ({g}_{\V A})^*
    =
    {d \over dt} \, {g}_{A \V}
    &=&
    2 g_{\V \V} \, A - 2 \, g_{A A} \, \Veff''
    + {\cal O} ( N^{-2} ) \,,
\\
    {d \over dt} \, {g}_{\V \V}
    &=&
    - 4 g_{\V \V} \, \V - 2 (g_{\V A} + g_{A \V}) \, \Veff''
    + {\cal O} ( N^{-2} ) \,,
\end{eqnarray}
\label{large-N-evolution-n=1}
\end {mathletters}
with initial conditions given by
$A(0) = a$, $\V(0) = \omega$, and
\begin{equation}
    \label{large-N-initial-conditions-n=1}
    \left(
	\matrix{
	g_{A A} &
	g_{A \V} \cr
	g_{\V A} &
	g_{\V \V} }
    \right)_{t=0}
    =
    {2\over N}
    \left(
	\matrix{ \phantom- a^2 & {i \over 2} \cr -{i \over 2} & \qtr \, a^{-2} }
    \right)
    + {\cal O}(N^{-2}) \,.
\end{equation}
 From Eq's.~(\ref{large-N-evolution-n=1}) and
(\ref{large-N-initial-conditions-n=1})
one may again see that to next-to-leading order,
the determinant of the variance matrix on the left-hand side of
(\ref{large-N-initial-conditions-n=1})
is a constant of the motion,
$\det g^{(2)}(t) = {\cal O} \left( N^{-3} \right)$.
To this order, our method gives exactly same predictions
as the Gaussian approximation of \cite{MDCBH}.
One may, of course, systematically extend the treatment to higher order
in $1/N$ simply by specializing the next-to-next-to-leading order results
in section~\ref {time-evolut}.

The evolution equations (\ref{large-N-evolution-n=1})
in this single-vector case may be cast in a more transparent form
by defining radial position and momentum operators via
\begin {equation}
    \hat A = \hat r^2 \,, \qquad
    \hat \V = \half ( \hat p \, \hat r^{-1} + \hat r^{-1} \, \hat p )
    \,,
\end {equation}
or equivalently
\begin {equation}
    \hat r = \hat A^{1/2} \,, \qquad
    \hat p = \hat A^{1/2} \, \hat \V - {\textstyle {i\over 2N}} \, \hat A^{-1/2}
    \,.
\end {equation}
These operators are canonically conjugate,
\begin {equation}
    i \, [ \hat p, \hat r] = 1/N \,,
\label {eq:[p,r]}
\end {equation}
and a short exercise rewriting the quantum equations of motion
(\ref {large-N-quantum-evolution}) yields
\begin {mathletters}
\begin{eqnarray}
    {d \over dt} \, \hat r &=& \hat p \,,
\\
    {d \over dt} \, \hat p &=& -\Ueff' (\hat r) \,,
\end{eqnarray}%
\label {eq:EOMpr}%
\end {mathletters}
where
\begin {eqnarray}
    \Ueff(r) &\equiv&
    \Veff(r^2) - {1 \over 8 N^2 \, r^2}
\nonumber\\
    &=&
    V(r^2) + {\textstyle {1 \over 8} \left( 1 - {3\over N} \right)
				    \left( 1 - {1\over N} \right) } \,
		r^{-2} \,,
\label {eq:Ueff}
\end {eqnarray}
and $\Ueff' = d\Ueff/dr$.
This is a well-known result:
$s$-wave dynamics in an $N$-dimensional central potential
is equivalent to one-dimensional quantum dynamics in
an effective radial potential $\Ueff$
containing an additional ``centrifugal'' potential
$
    {(N{-}3)(N{-}1) \over 8 N^2 \, r^2}
$
which is non-vanishing in all dimensions other than 1 and 3
\cite {YaffePhysToday,Witten}.
As seen in the commutation relations (\ref {eq:[p,r]}),
the parameter $1/N$ plays the role of $\hbar$
so that the large $N$ limit is precisely equivalent to
the semiclassical limit of ordinary one-dimensional quantum mechanics.

The next-to-leading order evolution equations (\ref {large-N-evolution-n=1})
for the coherent state expectation values and variances
of $A$ and $\Omega$
may be easily be converted to equivalent next-to-leading order
equations for expectations and variances of $p$ and $r$.
One finds,
\begin {mathletters}
\begin{eqnarray}
    {d \over dt} \, r
    &=&
    p + {\cal O}(N^{-2}) \,,
\\
    {d \over dt} \, p
    &=&
    - \Ueff' - \half g_{rr} \, \Ueff'''
    + {\cal O}(N^{-2}) \,,
\\
    {d \over dt} \, {g}_{rr}
    &=&
    g_{rp} + g_{pr}
    + {\cal O}(N^{-2}) \,,
\\
    {d \over dt} \, ({g}_{pr})^*
    =
    {d \over dt} \, {g}_{rp}
    &=&
    g_{pp} - g_{rr} \, \Ueff'' 
    + {\cal O}(N^{-2}) \,,
\\
    {d \over dt} \, {g}_{pp}
    &=&
    -(g_{rp} + g_{pr})  \, \Ueff''
    + {\cal O}(N^{-2}) \,.
\end{eqnarray}%
\label{large-N-evolution-rp}%
\end {mathletters}
Through next-to-leading order, these evolutions equations
are identical to the evolution equations (\ref {heisenberg-evolution})
for the usual semiclassical limit.%
\footnote
    {%
    This equivalence persists to all orders, of course,
    reflecting the exact correspondence between the
    operator equations of motion (\ref {eq:EOMxp})
    and (\ref {eq:EOMpr}).
    }
The initial variances differ, however,
due to the differing shapes of the initial
wavepackets (\ref {wf:QM}) and (\ref {eq:Psi}).
For our large $N$ coherent states,
\begin{equation}
    \left(
    \matrix{ g_{rr} & g_{rp} \cr g_{pr} & g_{pp} } \right)_{t=0}
    =
    {1\over 2N}
    \left(
	\matrix{ r^2 & pr {+} i~ \cr pr {-} i  & p^2 {+} r^{-2} }
    \right)
    + {\cal O}(N^{-2}) \,,
\end{equation}
[and once again $\det g^{(2)}(t) = {\cal O} \left( N^{-3} \right)$].
The form of this variance matrix
(including, for example,
the growth in the variance $g_{rr}$ with increasing $r$)
reflects the fact that the underlying $O(N)$ invariant coherent state
wavefunctions are not constant width one-dimensional Gaussians,
but rather $N$-dimensional Gaussians centered at the origin
with variable width.
Hence, the position of the peak in the resulting radial probability
distribution is positively correlated with the width
of the radial probability distribution about this peak.

For any given choice of the potential,
one may integrate the five equations (\ref {large-N-evolution-rp})
forward in time and obtain results which are accurate to ${\cal O}(N^{-2})$
[for times of order unity].
For better accuracy, one could extend the treatment to include
higher order correlations, as detailed in section~\ref {time-evolut}.

In light of the above exact correspondence
between the $O(N)$-invariant
dynamics of the single-vector model (\ref {eq:single-vector}),
and ordinary one-dimensional quantum dynamics in 
the the effective radial potential (\ref {eq:Ueff})
with $N$ playing the role of $\hbar$,
the previous discussion of stability of the truncated
moment equations in the semiclassical limit
immediately carries over to the large $N$ dynamics of
the single-vector model.
In particular, this means that one should
expect to see a decoherence time which scales as $N^{1/2}$,
beyond which truncations of the moment hierarchy
are no longer useful.
We have no reason to believe that the scaling of the decoherence
time with $N$ will be different in more general vector-like
large $N$ theories, such as the $\phi^4$ field theory (\ref {eq:Hlattice}),
as compared to the single-vector model.
Although we have no compelling proof to offer,
we expect that a decoherence time of order $N^{1/2}$
is a generic feature of vector-like large $N$ theories.%
\footnote
    {%
    It is interesting to note that,
    in contrast to the previous discussion
    of the semiclassical $\hbar\to0$ limit,
    examining $N$-dimensional free motion in the absence
    of any potential does not provide an example illustrating breakdown
    of the moment hierarchy based on $O(N)$ invariant coherent states.
    This is because the growth in the width of a spherically-symmetric
    Gaussian wavepacket is perfectly represented by a single one
    of the variable-width $O(N)$ invariant coherent states
    (\ref {eq:Psi}),
    unlike the earlier situation with fixed-width coherent states.
    Hence $O(N)$ invariant free motion is highly non-generic.
    For $O(N)$ invariant free motion (in the general case where
    $z$ is an $m \times m$ matrix and $\hat H_N = \half \tr \hat C$),
    one may show that the exact time evolution maps an initial
    coherent state $|z_0\rangle$ into another coherent state $|z(t)\rangle$
    with
    $
	z(t)^{-1} = z_0^{-1} + i t \, \id
    $.
    The operator equations of motion (\ref{eq:ABCeom}) may also be
    integrated exactly and show that
    $\hat C(t) = \hat C(0)$ is a constant of the motion, while 
    $\hat B = \hat B(0) + 2 \hat C t$, and
    $
	\hat A = \hat A(0) +
	\half \left[ \hat B(0) + \hat B(0)^T \right] t
	+ \hat C t^2
    $.
    This implies, for example, that for large time the variance
    $g^{AA}_{\alpha\beta,\mu\nu} \sim t^4/N$
    and so grows without bound.
    However, the mean value $\langle \hat A \rangle$ grows
    quadratically with $t$, and hence rms fluctuations remain
    bounded and of order $N^{-1/2}$ for all times.
    }

\section {Conclusions}

We have shown that a systematic hierarchy of time-local evolution
equations for a minimal set of equal-time correlation functions
may be derived in any theory having a classical (or large-$N$) limit
which fits within the general framework of section \ref {definitions}.
Truncating this hierarchy at the level of $k$'th order moments
({\em i.e.}, retaining up to $k$-point connected correlators)
yields results which are accurate up to order $1/N^k$.

However,
it is clear that the $t \to \infty$ and $\hbar \to 0$ (or $N \to \infty$)
limits are non-uniform.
At least in simple one degree of freedom (or single vector) models,
we have argued that
integrating the truncated moment evolution equations forward in time
yields results which, generically, cease to be a good approximation
to the true quantum dynamics beyond a decoherence time which scales
as $\hbar^{-1/2}$ (or $\sqrt N$).
The ordering of connected correlators which underlies the truncation
of the moment hierarchy is only valid for times small compared to
the decoherence time.
Going to higher orders in the truncation scheme
will not, in general, yield results which remain
valid for parametrically longer time intervals.

We expect, but have not demonstrated, that this $\sqrt N$ scaling of
the decoherence time is a general feature of large $N$ quantum dynamics.
It would obviously be worthwhile to investigate this further,
particularly in large $N$ models with many vectors.
In, for example, an $O(N)$ invariant lattice $\phi^4$ field theory,
it would clearly be desirable to understand the dependence of the
decoherence time on the energy of the initial state and the lattice volume.
If the $\sqrt N$ scaling of the decoherence time is generically true
this would, for example, imply that one cannot use truncated hierarchies
of large $N$ evolution equations (at least of the form considered here)
to study the non-equilibrium dynamics of
thermalization or hydrodynamic transport, as the relevant time
scales for these processes scale like $N$ in the large $N$ limit \cite {A&Y}.
We hope that future work will shed light on these issues.

\section* {Acknowledgment}

    One of the authors (LGY) wishes to thank Fred Cooper for
asking the question which stimulated this work.
A. Morozov was involved in initial portions of this investigation;
his efforts are gratefully acknowledged.


\begin{thebibliography}{99}

\bibitem {amit}
    D.~J.~Amit, 
    {\em Field Theory, the Renormalization Group, 
    and Critical Phenomena,}
    McGraw-Hill, 1978.

\bibitem {zinn-justin}
    J.~Zinn-Justin, 
    {\em Quantum Field Theory and Critical Phenomena,}
    Clarendon Press, Oxford, 1993.

\bibitem {Coleman}
    S.~Coleman,
    {\em Aspects of Symmetry,} {\em $1/N$},
    Cambridge University Press, 1985.

\bibitem {Yaffe}
    L.~G.~Yaffe,
    {\em Large $N$ limits as classical mechanics,}
    Rev.~Mod.~Phys.~{\bf 54}, 407--435 (1982).

\bibitem {Cooper1}
    F.~Cooper, S.~Habib, Y.~Kluger, E.~Mottola, J.~P.~Paz and P.~R.~Anderson,
    {\em Non-equilibrium quantum fields in the large N expansion,}
    hep-ph/9405352,
    Phys.~Rev.~D{\bf 50}, 2848--2869 (1994).

\bibitem {Cooper2}
    F.~Cooper, S.~Habib, Y.~Kluger and E.~Mottola,
    {\em Nonequilibrium dynamics of symmetry breaking in $\lambda \Phi^4$ 
    field theory,}
    hep-ph/9610345,
    Phys.~Rev.~D{\bf 55}, 6471--6503 (1997).

\bibitem {Boyanovsky1}
     D.~Boyanovsky, H.~J.~de Vega, R.~Holman, S.~P.~Kumar, 
     R.~D.~Pisarski and J.~Salgado,
     {\em Non-equilibrium real-time dynamics of quantum fields: 
     linear and non-linear relaxation in scalar and gauge theories,}
     hep-ph/9810209.

\bibitem {Boyanovsky2}
     D.~Boyanovsky, H.~J. de Vega and R.~Holman,
     {\em Non-equilibrium phase transitions in condensed matter 
     and cosmology: spinodal decomposition, condensates and defects,}
     hep-ph/9903534.

\bibitem {Boyanovsky3}
     D.~Boyanovsky, H.~J. de Vega, R.~Holman and J.~Salgado,
     {\em Non-equilibrium Bose-Einstein condensates, dynamical scaling 
     and symmetric evolution in large $N$ $\Phi^4$ theory,}
     hep-ph/9811273, Phys.~Rev.~D{\bf 59}, 125009 (1999).

\bibitem {Others}
    M.~D'Attanasio and T.~R.~Morris,
    {\em Large N and the renormalization group,}
    hep-th/9704094, Phys.~Lett.~{\bf B409}, 363--370 (1997).
    
\bibitem {critical-exponents-1}
    R.~Guida and J.~Zinn-Justin,
    {\em Critical exponents of the $N$-vector model,}
    cond-mat/9803240.
    
\bibitem {critical-exponents-2}
    V.~Yu.~Irkhin, A.~A.~Katanin and M.~I.~Katsnelson,
    {\em $1/N$ expansion for critical exponents of magnetic phase 
    transitions in $CP^{N-1}$ model at $2 < d < 4$,}
    cond-mat/9703011, Phys.~Rev.~B{\bf 54}, 11953--11956 (1996).

\bibitem {bose-cond}
    P.~Arnold and B.~Tomasik,
    {\em $T_c$ for dilute Bose gases: beyond leading order in $1/N$},
    cond-mat/0005197.

\bibitem {BW}
    L.~Bettencourt and C.~Wetterich,
    {\em Time evolution of correlation functions for classical and quantum
    anharmonic oscillators,}
    hep-ph/9805360.

\bibitem {BW2}
    G.~F.~Bonini and C.~Wetterich,
    {\em Time evolution of correlation functions and thermalization,}
    hep-ph/9907533, Phys.~Rev.~D{\bf 60}, 105026 (1999).

\bibitem {A&Y}
    P.~Arnold and L.~G.~Yaffe,
    {\em Effective theories for real-time correlations in hot plasmas,}
    hep-ph/9709449, Phys.~Rev.~D{\bf 57}, 1178--1192 (1998).

\bibitem {Arnold}
    See, for example, V.~I.~Arnold,
    {\em Geometrical methods in the theory of ordinary differential
    equations,}
    Springer, 1988.

\bibitem {MDCBH}
    B.~Mihaila, F.~Dawson, F.~Cooper, M.~Brewster and S.~Habib,
    {\em The quantum roll in d-dimensions and the large-d expansion,}
    hep-ph/9808234,
    and references therein.

\bibitem {LL}
    L.~Landau \& E.~Lifshitz,
    Vol I, {\em Mechanics,}
    Section 28,
    Pergamon Press, 3rd ed., 1996.

\bibitem {MP}
    L.~D.~Mlodinow and N.~Papanicolaou,
    {\em SO(2,1) Algebra and the large $N$ expansion in quantum mechanics,}
    Ann.~Physics {\bf 128}, 314--334 (1980).

\bibitem {YaffePhysToday}
    L.~G.~Yaffe,
    {\em Large-$N$ quantum mechanics and classical limits},
    Phys.~Today {\bf 36}, no. 8, 50--57, (1983).

\bibitem {Witten}
    E.~Witten,
    in {\em Proc. of Cargese Summer Inst. on Quarks and Leptons},
    M.~Levy, ed., Plenum, NY (1980).

\end{thebibliography}
\end{document}